\documentclass[11pt,draftcls,onecolumn]{IEEEtran} 

\usepackage{fullpage,amsmath,amssymb,amsthm,setspace,citesort}
\usepackage{epsfig,graphicx,subfigure,wrapfig,psfrag}		

\graphicspath{{./bio/}{./figs/}}

\def\reals{\mathbb{R}}
\def\comps{\mathbb{C}}

\newcommand{\Reo}[1]{\operatorname{Re}\left\{ #1 \right\}}
\newcommand{\Imo}[1]{\operatorname{Im}\left\{ #1 \right\}}

\newtheorem{definition}{Definition}[section]
\newtheorem{lemma}{Lemma}[section]
\newtheorem{thm}{Theorem}[section]
\newtheorem{cor}{Corollary}[section]

\begin{document}

\title{Stable Takens' Embeddings for\\Linear Dynamical Systems}
\date{}
\author{Han Lun Yap and Christopher J. Rozell*\thanks{*Corresponding Author. Manuscript received October 2010. 
This work was partially supported by NSF grant CCF-0830456 and DSO National Laboratories, Singapore. 
Copyright (c) 2011 IEEE. Personal use of this material is permitted. However, permission to use this material for any other purposes must be obtained from the IEEE by sending a request to pubs-permissions@ieee.org.

The authors are with the School of Electrical and Computer Engineering at the Georgia Institute of Technology.  Preliminary versions of portions of this work appeared in~\cite{yap2010stabletakens}.}
}
\maketitle
\begin{abstract}

Takens' Embedding Theorem remarkably established that concatenating $M$ previous outputs of a dynamical system into a vector (called a \emph{delay coordinate map}) can be a one-to-one mapping of a low-dimensional attractor from the system state space. 
However, Takens' theorem is fragile in the sense that even small imperfections can induce arbitrarily large errors in this attractor representation.  We extend Takens' result to establish deterministic, explicit and non-asymptotic sufficient conditions for a delay coordinate map to form a \emph{stable embedding} in the restricted case of linear dynamical systems and observation functions.  
Our work is inspired by the field of Compressive Sensing (CS), where results guarantee that low-dimensional signal families can be robustly reconstructed if they are stably embedded by a measurement operator.  
However, in contrast to typical CS results, i) our sufficient conditions are independent of the size of the ambient state space, and ii) some system and measurement pairs have fundamental limits on the conditioning of the embedding (i.e., how close it is to an isometry), meaning that further measurements beyond some point add no further significant value.
We use several simple simulations to explore the conditions of the main results, including the tightness of the bounds and the convergence speed of the stable embedding. We also present an example task of estimating the attractor dimension from time-series data to highlight the value of stable embeddings over traditional Takens' embeddings.

\end{abstract}


\section{Introduction}

Of the many types of data confronting signal processing researchers, time series data is perhaps one of the most common.  While there are many possible ways to analyze a time series, one of the most important tasks in many areas of science and engineering is to characterize (or predict) the state of a dynamical system from a stream of its output data~\cite{book_brockwell2002timeseries,kantz2004nonlinear}.  This type of state identification can be particularly challenging because the internal (possibly high-dimensional) system state $x(t) \in \reals^N$ is often only indirectly observed via a one-dimensional time series of measurements produced through an observation function $s(t) = h(x(t))$, where $h:\reals^N\to\reals$.

Surprisingly, when the dynamical system has low-dimensional structure because the state is confined to an attractor $\mathcal{M}$ of dimension $d$ ($d<N$) in the state space, Takens' Embedding Theorem~\cite{takens,embedology} shows that complete information about the hidden state of this system can be preserved in the time series output data $s(t)$.  Indeed, many systems of interest do have this type of structure~\cite{book_strogatz1994nonlinear}, and a variety of algorithms for tasks such as time series prediction and attractor dimension estimation exploit Takens' result~\cite{kantz2004nonlinear}.  Specifically, Takens defined the \emph{delay coordinate map} $F: \reals^N \rightarrow \reals^M$ as a mapping of the state vector $x(t)$ to a point in the \emph{reconstruction space} $(\reals^M)$ by taking $M$ uniformly spaced samples of the past time series  (with sampling interval $T_s$) and concatenating them into a single vector, 
\begin{equation}
	F(x(t)) = [s(t) \; s(t-T_s) \; s(t-2T_s) \;\cdots \; s(t-(M-1)T_s)]^T.	
	\label{eqn:dcm}
\end{equation}
Takens' main result \cite{takens} (later refined in \cite{embedology}) states that (under a few conditions on $T_s$ discussed later) for almost every smooth observation function $h(\cdot)$, the delay coordinate map is an \textit{embedding}\footnote{An \textit{embedding} is a \textit{one-to-one immersion}.} of the state space attractor $\mathcal{M}$ when $M>2d$.  In other words, despite the state being hidden from direct observation, the topology of the attractor that characterizes the dynamical system can be preserved in the time series data when it is arranged into a delay coordinate map.

In the absence of imperfections such as measurement or system noise, Takens' result indicates that a delay coordinate map should be as useful for characterizing a system as direct observation of the hidden system state.  
However, in the presence of noise, a one-to-one mapping may not be sufficient to guarantee the robustness of any processing  performed in the reconstruction space (e.g., dimensionality estimation).  
The main underlying problem is that while Takens' theorem guarantees the preservation of the attractor's \emph{topology}, it does not guarantee that the \emph{geometry} of the attractor is also preserved.  
For example, Takens' result guarantees that two points on the attractor $\mathcal{M}$ do not map to the same point in the reconstruction space, but there are no guarantees that close points on the attractor remain close under this mapping (or far away points remain far away).  Consequently, relatively small imperfections could have arbitrarily large effects when the delay coordinate map is used in applications.

In the signal processing community, recent work has highlighted the importance of well-conditioned measurement operators to ensure the geometry of a low-dimensional signal family is preserved.  
Consider a signal class $\widetilde{\mathcal{M}}$ with intrinsic dimension $d$ residing in $\reals^N$ and measurement operator $\widetilde{F}: \reals^N \rightarrow \reals^M$.
We call $\widetilde{F}$ to be a \emph{stable embedding} of the signal class $\widetilde{\mathcal{M}}$ if
for all distinct pairs of points $x,y \in \widetilde{\mathcal{M}}$ their pairwise distances are preserved by satisfying
\begin{eqnarray}
	C(1- \delta) \le \frac{\|\widetilde{F}(x)-\widetilde{F}(y)\|_2^2}{\|x - y\|_2^2} \le C(1 + \delta).
\label{eqn:cond_embed}
\end{eqnarray}
The \emph{scaling constant} $C$ could be absorbed into $\widetilde{F}$ and the \emph{conditioning number} $0\leq \delta<1$ bounds how much pairwise distances between signals in  $\widetilde{\mathcal{M}}$ can change when mapped by $\widetilde{F}$ (i.e., how near $\widetilde{F}$ is to an isometry). 
The Johnson-Lindenstrauss (JL) lemma \cite{dasgupta2002elementary,achlioptas2003database} gives an example of a stable embedding of a signal class $\widetilde{\mathcal{M}}$ consisting of a point cloud of $d=|\widetilde{\mathcal{M}}|$ distinct points in $\reals^N$. 
In this result, a random measurement matrix $\widetilde{F}$ with $M=O(\log(d))$ rows ensures that \eqref{eqn:cond_embed} holds with high probability for all pairs of points in the point cloud $\widetilde{\mathcal{M}}$.
Another example is the recent work in the field of \emph{compressed sensing} (CS)~\cite{CompSampCand,CompSenDon}, %
where the canonical results show that similar random matrices $\widetilde{F}$ satisfy the \emph{Restricted Isometry Property} (RIP) with high probability when $M=O(d\log(N/d))$~\cite{jlcs,mendelson2008uniform}.  
The RIP guarantees that~\eqref{eqn:cond_embed} holds for all pairs of $d$-sparse signals (i.e., the signal family $\widetilde{\mathcal{M}}$ is comprised of signals on the union of all $d$-dimensional subspaces  within $\reals^N$).  
Beyond extending the concept of the JL lemma from a finite point cloud to an infinite signal family, the CS results  show the value of stable measurement operators by also making  guarantees about efficient and robust signal recovery from these measurements.  The notion of a stable embedding has also been extended to other signal models~\cite{baraniuk2008model}, including manifold signal families~\cite{wakin_embedding,clarkson2008tighter}. 
The latter can be seen as an extension of Whitney's Embedding Theorem~\cite{whitney}; while Whitney's Embedding Theorem ensures a one-to-one mapping of a manifold $\widetilde{\mathcal{M}}$ with dimension $d$ for almost any smooth projection function $\widetilde{F}$ given that $M > 2d$, the results in~\cite{wakin_embedding} further guarantee that~\eqref{eqn:cond_embed} holds over this signal family for a given $\delta$ with high probability when $M=O(d\log(N))$ and $\widetilde{F}$ is a random orthoprojector.\footnote{The required number of measurements $M$ in~\cite{wakin_embedding} also depends on some properties of the manifold (e.g., the maximum curvature). Clarkson \cite{clarkson2008tighter} later improved upon $M$ to remove the dependence on the ambient dimension $N$ and reduce the dependence on certain properties of the manifold.}

While the notion of embedding the state of a dynamical system may seem far removed from the CS results, there is actually a close connection.  It is well-known that Takens' Embedding Theorem can be viewed as a special case of Whitney's Embedding Theorem where the measurement operator $\widetilde{F}$ is restricted to forming a delay coordinate map (i.e., $\widetilde{F}=F$) and $\widetilde{\mathcal{M}}$ is taken to be the state space attractor (i.e., $\widetilde{\mathcal{M}}=\mathcal{M}$)~\cite{kantz2004nonlinear}.  The main contribution of this paper is to further these connections by establishing sufficient conditions whereby the delay coordinate map is a stable embedding of the state space attractor for linear systems with linear observations functions.
Indeed, the main technical result of this paper establishes deterministic, explicit and non-asymptotic sufficient conditions for the delay coordinate map to be a stable embedding with a given conditioning $\delta$.
We also explore the meaning of these conditions for characterizing systems via delay coordinate maps.  
In particular, the results of this exploration are interesting because they contrast with the standard CS results in two principle ways: $(i)$ the conditioning of the operator cannot always be improved by taking more measurements, as some system/observation pairs will have a fundamental limit in how well the system geometry can be preserved, and $(ii)$ the necessary number of measurements scales with the dimension of the attractor $d$ but is independent of the dimension of the ambient space $N$.

Due to the importance of nonlinear systems, a similar general stable embedding result for nonlinear dynamical systems is obviously of great interest.  Linear systems have a wealth of tools available for their analysis and the language of ``attractors'' is uncommon when studying these relatively simple systems (despite the notion of an attractor being technically well-posed for the restricted class of linear systems we study here).  Therefore, beyond just contributing a new tool for linear systems analysis and design (as demonstrated in the example of Section~\ref{sec:dimest}), our present results are perhaps most valuable for elucidating some of the unique issues that arise when trying to stabilize the embeddings of dynamical systems, helping to pave the way for extensions to nonlinear systems.

%

\section{Background and Related Work}
\label{sec:prelim}

In this section we will briefly review some preliminaries, including a precise statement of Takens' theorem, attractors of linear systems, and related work in stable embeddings of attractors and manifolds.

\subsection{Linear Systems and Delay Coordinate Maps}
\label{sec:linear_sys}

Let a dynamical system be defined by the differential equation: 
\begin{eqnarray}
	\dot{x}(t) = \Psi\left(x(t) \right), 
	\label{eq:diff_eq}
\end{eqnarray}
where $x(t)\in\reals^N$ is the system state at time $t$, and $\Psi: \reals^N \rightarrow \reals^N$ is a smooth function.
As stated earlier, in this paper we will restrict our examination to embeddings of linear dynamical systems where $\Psi \in \reals^{N \times N}$ is a matrix.  
Before going on, our discussion of these systems will require us to establish a basic notation for complex vector spaces. For $u = [u_1 \; \cdots \; u_N]^T \in \comps^N$, we denote the complex variable by $j$, the (element-wise) complex conjugate by ${u}^{*}$ and the Hermitian transpose by $u^H=(u^*)^T$.

Given the system matrix $\Psi$ and the definition of a dynamical system~\eqref{eq:diff_eq}, knowing the state at some fixed time $t_0$ is equivalent to knowing the path that the system takes to and from that state (called the \emph{flow}).  Classic results in linear systems theory~\cite{brogan_book} show that the explicit solution for this path is given by a matrix multiplication: $x(t_0+t) = e^{\Psi t} x(t_0) = \Phi_t x(t_0)$, where $\Phi_t =e^{\Psi t}$ is the \emph{flow matrix}.  Note that this solution is valid for positive or negative values of $t$, describing the flow both forward and backward from time $t_0$.

Delay coordinate maps that embed points on the attractor of a dynamical system are intimately connected with the flow of the system approaching that point.  
In particular, forming a delay coordinate map of a specific point in the state space requires collecting samples of the system flow backward in time from that point at regular intervals $T_s$.  To enable mathematical descriptions of this sampling operation along the flow, we suppress the implicit dependence on the sampling time $T_s$ and define the compact notation for the flow matrix as $\Phi = \Phi_{-T_s}$ so that $x(t-T_s)= \Phi x(t)$.  The delay coordinate map $F$ with $M$ delays given in~\eqref{eqn:dcm} for the case of linear dynamical systems and linear observation functions $h \in \reals^N$ can then be written as a $M \times N$ matrix:
\begin{equation}
	F = \left(h \;|\; \Phi^T h \;|\; \cdots \;|\; (\Phi^{M-1})^T h \right)^T.
	\label{eq:DCM_linear_arb_h}
\end{equation}

To ensure that the linear dynamical systems under consideration have non-trivial steady-state behavior (i.e., oscillations rather than convergence to a fixed point), we restrict our study to the class of systems ${\mathcal{A}(d)}$ described in the following definition.
\begin{definition} 
	\label{def:A-eigenvalues}
	We say that a linear dynamical system in $\reals^N$ defined by~\eqref{eq:diff_eq} is of \textbf{Class $\mathbf{\mathcal{A}(d)}$} for $d \le \frac{N}{2}$ if the system matrix $\Psi$ is real, full rank and has distinct eigenvalues. Moreover, $\Psi$ has only $d$ strictly imaginary\footnote{A number $x$ is \textit{strictly imaginary} if $\operatorname{Re}\{x\} = 0$.  This condition ensures that the system modes corresponding to these eigenvectors have persistent oscillation in the steady-state response.} conjugate \emph{pairs} of eigenvalues and the rest of its eigenvalues have real components strictly less than 0. The strictly imaginary conjugate pairs of eigenvalues are called the \textbf{$\mathcal{A}$-eigenvalues} and they can be expressed as $\{\pm j \theta_i\}_{i=1}^{d}$ where $\theta_1, \cdots, \theta_d > 0$ are $d$ distinct numbers. The corresponding unit-norm \textbf{$\mathcal{A}$-eigenvectors} are $v_1, {v_1}^{*}, \cdots, v_d, {v_d}^{*}$.	  The corresponding eigenvalues of the flow matrix $\Phi$ are called the  \textbf{$\mathcal{A}_{\Phi}$-eigenvalues}, and are given by $\{ e^{\pm j \theta_i T_s} \}_{i=1}^{d}$.
\end{definition}
\noindent Furthermore, we define $\Lambda = \mathrm{diag}\left(j\theta_1, -j\theta_1, \dots, j\theta_d, -j\theta_d\right)$
as the diagonal matrix composed of the $\mathcal{A}$-eigenvalues and $V = \left( v_1\;|\; {v_1}^{*} \;|\; \cdots \;|\; v_d \;|\; {v_d}^{*}  \right) \in \comps^{N \times 2d}$
as the concatenation of the $\mathcal{A}$-eigenvectors into a matrix with $\operatorname{rank}(V) = 2d$.
Since $\Phi$ is the matrix exponential of $\Psi$, it is well-known that they share the same eigenvectors~\cite{moon2000mathematical}.  
Therefore, if we denote $D = D_{-T_s} = e^{- \Lambda T_s}$ as the diagonal matrix comprised of the $\mathcal{A}_{\Phi}$-eigenvalues, then we have $\Phi V = V D$.

In order to have a meaningful notion of an embedding, the dynamical system must have its state trajectory confined to a low-dimensional attractor in the state space.  Even if the system has transient characteristics from a given starting point, the embedding of a system is only considered in steady-state when these transients have disappeared.
Considering the steady-state dynamics of the system, we make explicit the notion of an \textit{attractor} through the following definition.
\begin{definition}
	\label{def:mathcal_M}
	Let a linear dynamical system be of class $\mathcal{A}(d)$ and let $x_0 = V \alpha_0 \in \reals^N$ for some $\alpha_0 \in \comps^{2d}$ be an arbitrary initial state of the system.\footnote{We only need to consider $x_0$ in the span of the columns of $V$ because any orthogonal components vanish in steady-state.}  We define the \textbf{attractor} of this linear dynamical system  to be $\mathcal{M} = \left\{ x \in \reals^N \;|\; x = V e^{\Lambda t} \alpha_0\;,\; t \in \reals \right\}$.
\end{definition}

It is easy to see that $\mathcal{M}$ lives in the span of $V$. Also, the attractor of the system clearly depends on the initial state of the system.  
Because the main results of this paper do not depend on the choice of initial state,
we will simply refer to the fixed attractor as  $\mathcal{M}$ and suppress the implicit dependence on the initial state.
Additionally, one can check that this definition meets the fundamental notion of an attractor, i.e., that any point on the attractor $\mathcal{M}$ when projected backwards (or forward) in time by $\Phi$ will remain on $\mathcal{M}$. Specifically, for any $x\in\mathcal{M}$, we can write $x = V \alpha_x$, where $\alpha_x = e^{\Lambda t_x} \alpha_0$ for some $t_x \in \reals$.  Then we see that for some $D$ (the diagonal matrix comprised of the $\mathcal{A}_{\Phi}$-eigenvalues as defined earlier) and any $k \in \mathbb{Z}$,  $\Phi^k x = \Phi^k V \alpha_x = V D^k \alpha_x$, meaning that $x$ remains on the attractor even when it is projected forward or backward in time. Finally, while we will not show this in detail due to space constraints, one can show that for each $i$ the state $x(t)$ is moving in an elliptical orbit on the span of $\Reo{v_i}$ and $\Imo{v_i}$ 
with angular speed proportional to $\theta_i T_s$.

\begin{figure*}
	\hfil
	\begin{minipage}[t]{0.4\linewidth}
	\centerline{\epsfysize = 45mm \epsffile{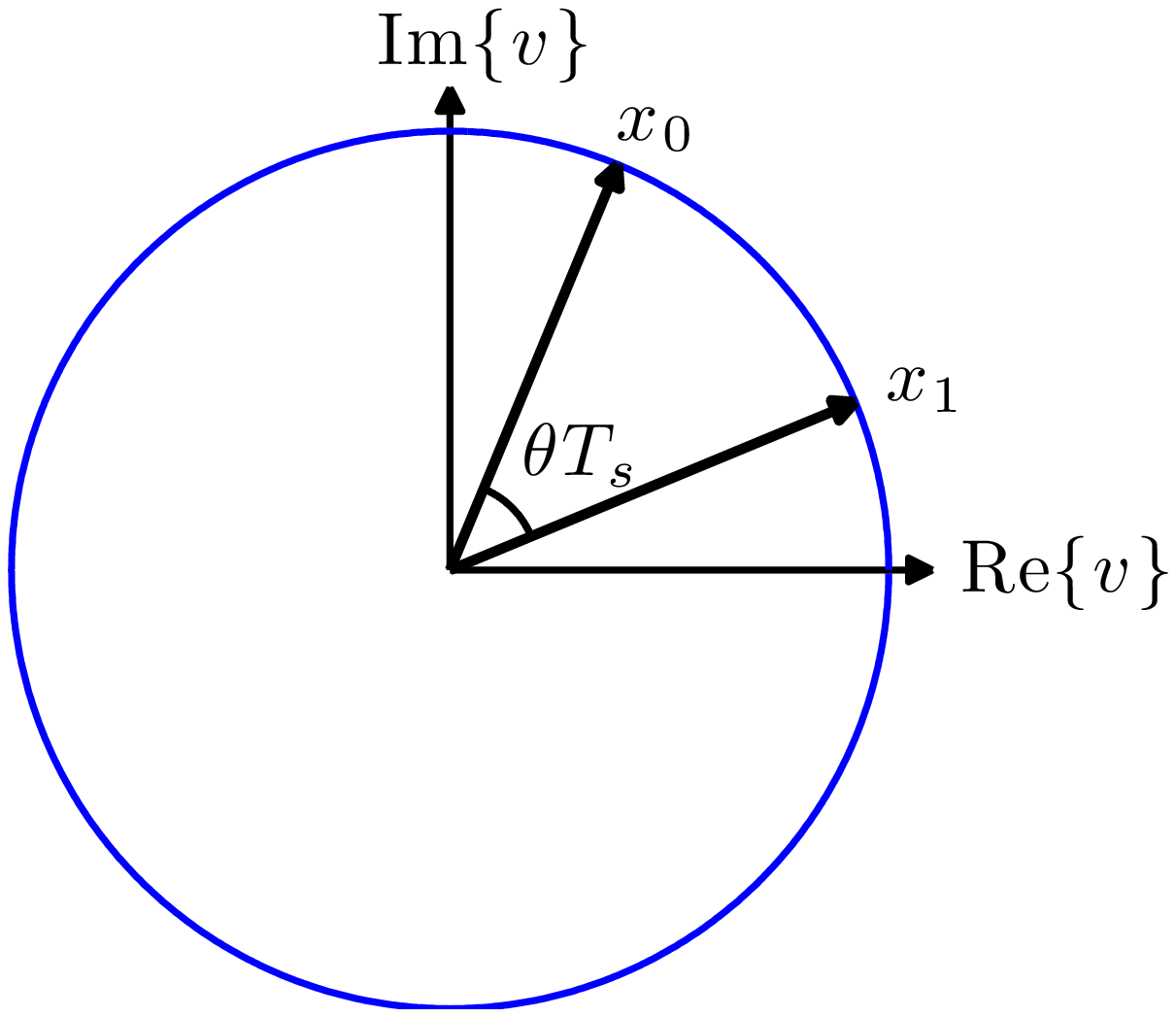}}
	\vspace{-4mm}
	\centerline{\small$\quad$~(a)}
	\end{minipage}
	\hfil
	\begin{minipage}[t]{0.4\linewidth}
	\centerline{\epsfysize = 45mm \epsffile{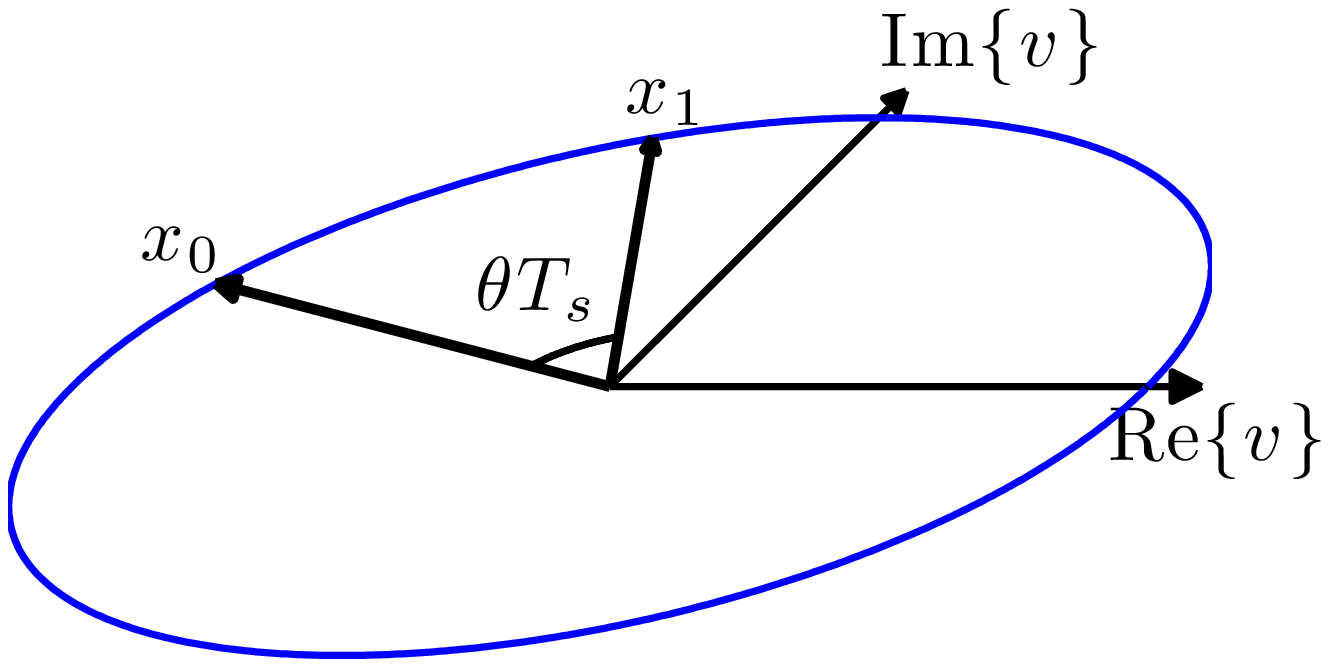}}
	\vspace{-4mm}
	\centerline{\small$\quad$~(b)}
	\end{minipage}
\vspace{-2mm}
\caption{\sl\small Examples of attractors of linear dynamical systems of class $\mathcal{A}(d)$ in $\reals^N$ for $N = 2$ and $d = 1$ with sampling interval $T_s = 1$. (a) A system attractor when $\theta = \frac{\pi}{4}$ and $v = \frac{1}{\sqrt{2}}[1,\; j]^T$. This results in a circular attractor where the system progresses at an angular speed determined by $\theta$. (b) A system attractor when $\theta = \frac{\pi}{4}$ and $v = [0.8165 + 0.4082j, \; -0.4082j]^T$. Here the system also progresses at the same angular speed, but the attractor is now an ellipse.}
\label{fig:attractor_example}
\vspace{-8mm}
\end{figure*}

For clarity and to build intuition, we give two brief examples where $N = 2$, $d = 1$ and  $T_s = 1$.  For the first example, consider a dynamical system of class $\mathcal{A}(d)$ with $\mathcal{A}$-eigenvalue $\theta = \frac{\pi}{4}$ and $\mathcal{A}$-eigenvector $v = \frac{1}{\sqrt{2}}[1,\; j]^T$.  Shown in Figure~\ref{fig:attractor_example}(a) is the resulting circular attractor of this system, along with the real and imaginary components of the $\mathcal{A}$-eigenvector and a pair of states separated in time by $T_s$ (which corresponds to a separation of $\theta T_s$ in angle).  For the second example, consider 
a dynamical system of class $\mathcal{A}(d)$ with the same parameters except that the $\mathcal{A}$-eigenvector is now defined as $v = [0.8165 + 0.4082j, \; -0.4082j]^T$.
Shown in Figure~\ref{fig:attractor_example}(b) is the resulting elliptical attractor and state time samples, illustrating that the angular speed is unchanged at $\theta T_s$.
In both of these examples, the elongation of the ellipse is determined by the inner product between $\Reo{v}$ and $\Imo{v}$, which governs how well the attractors fill the dimensions of the state space that it occupies.  While this is intuitive to visualize in the present case of $d=1$, for general $d > 1$ this elongation is determined by the ratio between
the smallest and largest eigenvalues of $V^H V$, denoted $A_1$ and $A_2$, respectively.  When $A_1 = A_2$, the system state revolves around a circle when projected onto each of the subspaces spanned by $\Reo{v_p}$ and $\Imo{v_p}$ for $p = 1, \cdots, d$, and the resulting attractor is a product of these circular orbits.  However when $A_2 \gg A_1$, the projection of the attractor onto some (or all) of these subspaces will be a highly elongated ellipse, therefore not equally filling the dimensions of the state space that it occupies.

\subsection{Attractor Embeddings}
\label{sec:Takens}

The following theorem is an extension of Takens' original result \cite{takens}, and gives a lower bound on the number of measurements $M$ sufficient to ensure that a delay coordinate map $F$ defined as in \eqref{eqn:dcm} is a one-to-one mapping from the state space attractor to the measurement (reconstruction) space.
\begin{thm}[Takens' Embedding Theorem~\cite{embedology}]
	\it
	\label{thm:takens}
	Assume the dynamical system converges to an attractor $\mathcal{M}$ of dimension $d$ and pick a sampling interval $T_s > 0$. 
	Let $M > 2d$ and suppose $\mathcal{M}$ has a finite number of equilibria, no periodic orbits of $\Psi$ of period $T_s$ or $2T_s$, and at most finitely many periodic orbits of period $kT_s$ for $k = 3, \cdots, M$.
	Then for almost every smooth function $h$, the {delay-coordinate map} $F$ is one-to-one on $\mathcal{M}$.
\end{thm}
\noindent The notion of ``almost every'' used in the theorem above is technical (see~\cite{embedology} for details), but is consistent with the heuristic notion that out of all possible functions $h$, most will indeed work.

In this paper we consider the question of when the one-to-one property described in Theorem~\ref{thm:takens} can be improved to become a stable embedding where $F$ is (nearly) an isometry that preserves the geometry of $\mathcal{M}$.  Specifically, we introduce the following definition to formalize the notion of a stable embedding.
\begin{definition}
Suppose we have a dynamical system in $\reals^N$ that converges to an attractor $\mathcal{M}$ and a linear map $F: \reals^N \rightarrow \reals^M$.
We say that $F$ is a \emph{stable embedding} of $\mathcal{M}$ with \emph{conditioning} $\delta$ if for all $x,y\in\mathcal{M}$ and for some \emph{scaling constant} $C$, we have
\begin{equation}
	C(1- \delta) \le \frac{\|F(x)-F(y)\|_2^2}{\|x - y\|_2^2} \le C(1 + \delta).
\label{eqn:ARIP}
\end{equation}
\end{definition} 
\noindent Note that smaller values of $\delta$ in the above definition imply a more stable embedding because it guarantees that the map is closer to an isometry. We also note that preservation of Euclidean distances also implies that the geodesic distances between points on the attractor are preserved~\cite{wakin_embedding}.
Because Taken's result only tells us that the delay coordinate map $F$ is a one-to-one mapping, it does not guarantee any specific value of the conditioning, meaning that $\delta$ could be arbitrarily close to 1 and the embedding could be highly unstable.

\begin{figure*}
\begin{center} 
	\hfil
	\begin{minipage}[t]{0.3\linewidth}
		\centerline{\epsfysize = 40mm \epsffile{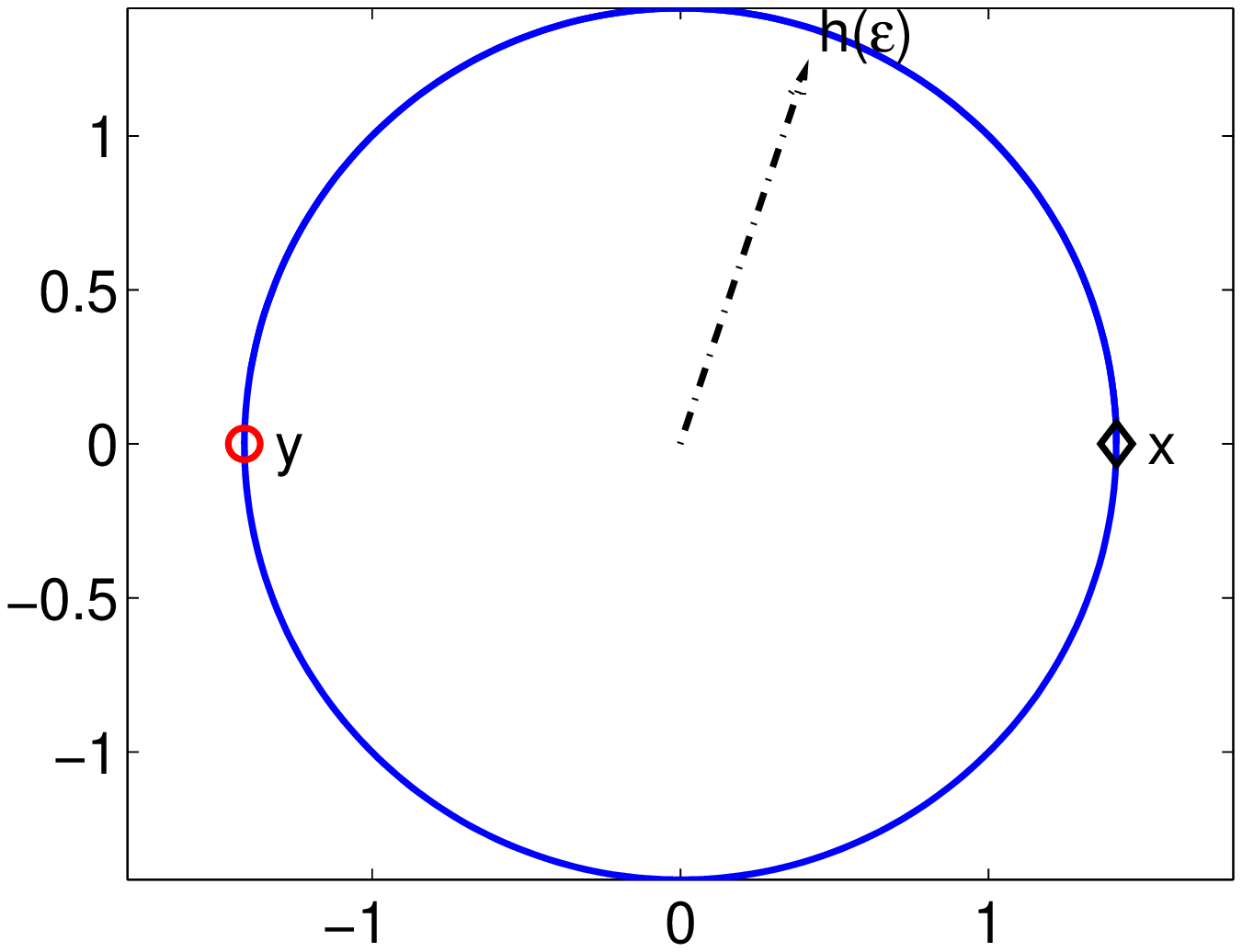}}
		\vspace{-1mm}
		\centerline{\small$\quad$~(a)}
	\end{minipage}
	\hfil
	\begin{minipage}[t]{0.3\linewidth}
		\centerline{\epsfysize = 40mm \epsffile{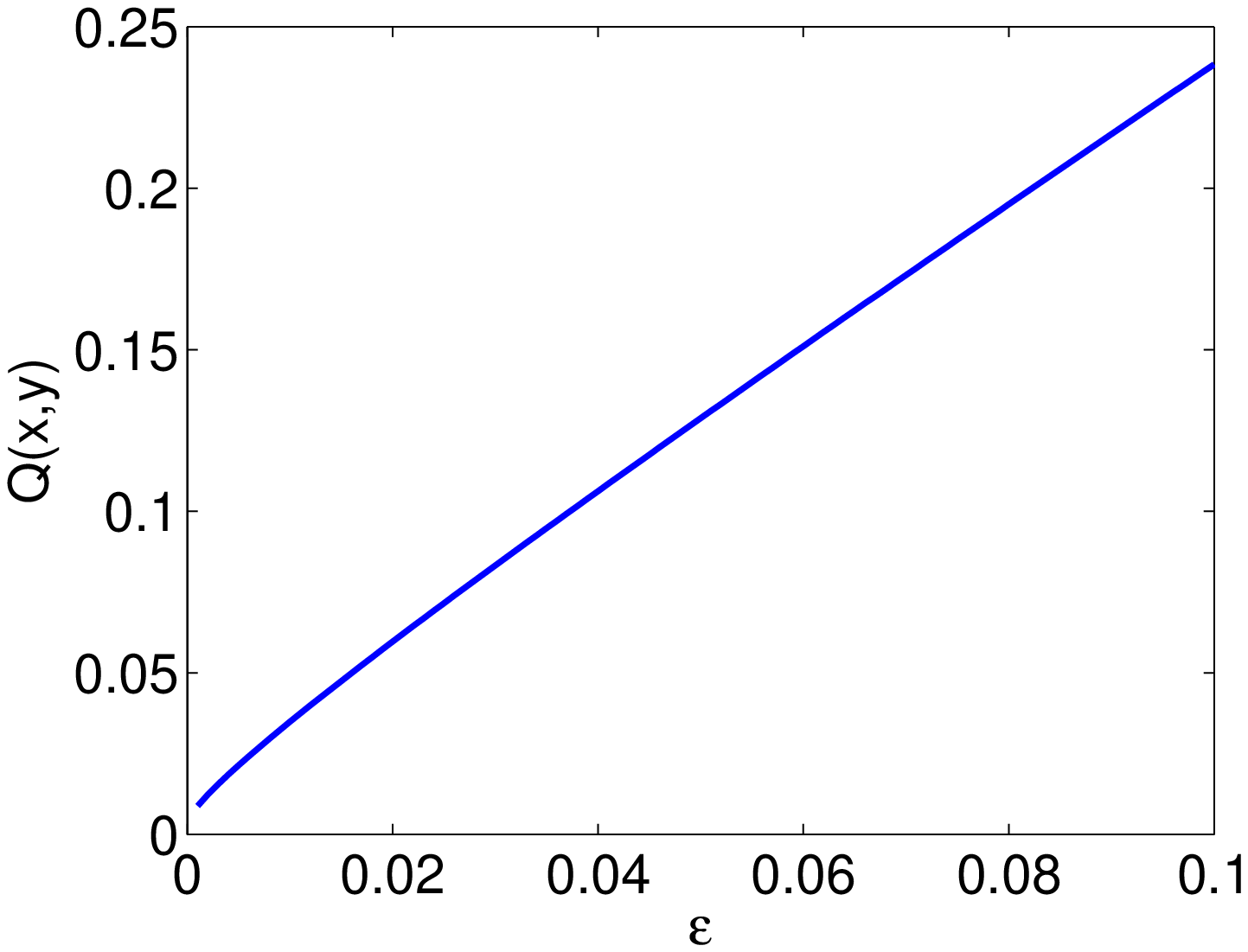}}
		\vspace{-1mm}
		\centerline{\small$\quad$~(b)}
	\end{minipage}
	\hfil
	\begin{minipage}[t]{0.3\linewidth}
		\centerline{\epsfysize = 40mm \epsffile{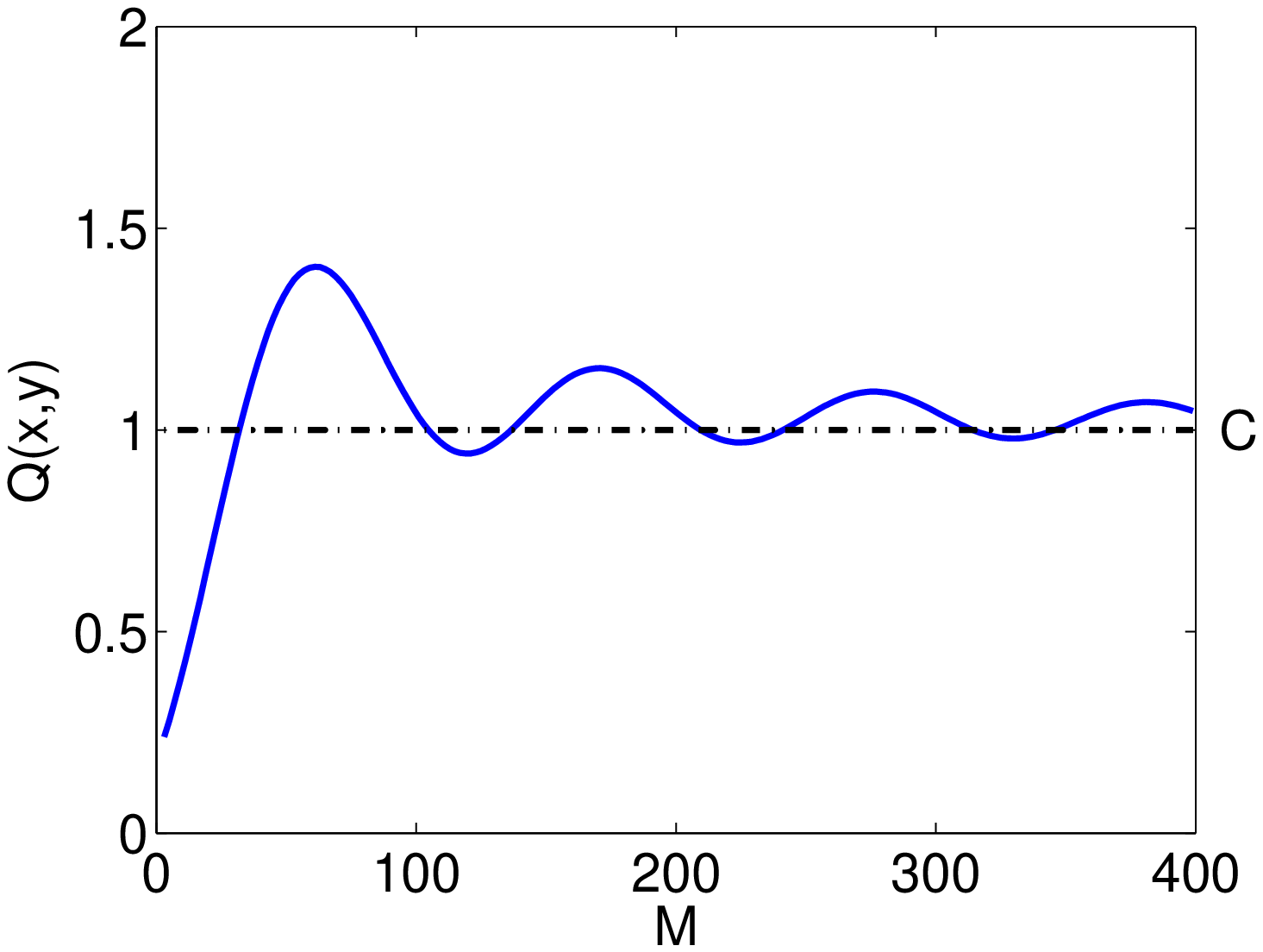}}
		\vspace{-1mm}
		\centerline{\small$\quad$~(c)}
	\end{minipage}
\end{center}
\vspace{-3mm}
\caption{\sl\small Examining the conditioning of Takens' embeddings.
(a) The large (blue) circle shows the attractor of the linear system. The (black) diamond and (red) circle markers show 2 different points $x,y$ that we pick on the opposite ends of the attractor. The arrow depicts the measurement function $h(\epsilon)$.
(b) The graph shows $Q(x,y)$ for the points $x,y$ in Figure \ref{fig:lead1_Q}(a) over a range of values of $\epsilon$ from 0.01 to 0.1. The number of measurements $M$ is fixed at 3, the minimum required by Takens' theorem. 
(c) Here $Q(x,y)$ is plotted for $M$ ranging from 3 to 400 (with $\epsilon$ fixed at 0.1), suggesting a near isometry for $F$ as $M$ increases.} 
\label{fig:lead1_Q}
\vspace{-8mm}
\end{figure*}

To see why Takens' Embedding can be insufficient, we present an illustrative example where the conditioning of the embedding can be made arbitrarily bad when  $M$ is the minimum number of delays  necessary to satisfy the sufficient conditions of Theorem~\ref{thm:takens}.  Consider a linear system of class $\mathcal{A}(1)$ with $N=2$, $T_s=1$,  $\mathcal{A}$-eigenvalue $\theta = 0.03$ and $\mathcal{A}$-eigenvector $v = \frac{1}{\sqrt{2}}[1, \; j]^T$. This system has a circular attractor as depicted in Figure \ref{fig:lead1_Q}(a).  We set the observation function to be $h = \sqrt{\frac{2}{M}}[\sqrt{\epsilon},\; \sqrt{1-\epsilon}]^T$.\footnote{As will be described in Theorem \ref{thm:stability_thm}, the observation function is normalized  so that we have scaling constant of $C=1$ regardless of $M$.}   Given a particular pair of points $x,y$ on opposite ends of the circular attractor (shown in Figure \ref{fig:lead1_Q}(a)), we examine the ratio $Q(x,y) = \frac{\|F(x) - F(y)\|_2^2}{\|x - y\|_2^2}$, where $F$ is the delay coordinate map given in \eqref{eq:DCM_linear_arb_h}.  Note that if $F$ is a perfect isometry then $Q(x,y)=1$, and  we must have $Q(x,y)>0$ for $F$ to be one-to-one.  
Fixing the number of measurements at $M=3$ (the minimum required by Takens' theorem), Figure \ref{fig:lead1_Q}(b) shows the behavior of $Q(x,y)$ for this pair of points as a function of $\epsilon$.  We see that while meeting the sufficient conditions of Takens' Theorem, 
$\lim_{\epsilon\to 0} Q(x,y)=0$.  Stated another way, by adjusting the parameter $\epsilon$ the conditioning of $F$ can be made arbitrarily bad for this pair of points.  To see that this is not simply a bad pairing of the measurement function to the system, note that for any admissible choice of $h$ there would exist a pair of points that would behave the same way.\footnote{One can imagine this by rotating the points $x,y$ by an angle equivalent to the angle between the new measurement function and the given $h$.}  To explore this example further, Figure \ref{fig:lead1_Q}(c) plots $Q(x,y)$ with $\epsilon = 0.1$ and varying $M$ from 3 to 400. We see that with increasing $M$, the ratio $Q(x,y)$ increases, oscillates and converges to a value of $C = 1$. 
This provides evidence suggesting that as $M$ increases, the conditioning of $F$ improves because the distance between this pair of points is preserved with increasing fidelity.
This effect is not predicted by Theorem~\ref{thm:takens}, but will be shown in our main results in Section~\ref{sec:STE_Ad}.

\subsection{Related Work}

Independently but at nearly the same time as Takens' original work, 
Aeyels \cite{aeyels1981generic} looked at the same problem from a control theory standpoint. He showed that the delay-coordinate map is related to the observability criteria and that given any system in $N$ dimensions (not just one confined to an attractor), a generic choice of observation function $h$ guarantees that the system is observable as long as  $M\geq 2N + 1$. 
Similar to the idea of a stable embedding, the authors in~\cite{Kang2009a} developed a robustness measure for the observability of dynamical systems. 
Stated in the language of delay coordinate maps and sampled systems, they defined a system as \emph{observable with precision} $(\epsilon, \delta)$ if for any two states $x,y$ on a trajectory in the state space, $\|F(x) - F(y)\|_2 \le \epsilon$ implies $\|x - y\|_2 \le \delta$. 
In addition to Takens' original investigation of attractor embeddings~\cite{takens}, significant advances were made by Sauer et al. \cite{embedology} to extend these results to include attractors of non-integer dimensions (i.e., strange attractors) and to make the definition of ``almost every'' more in line with notions of an event that occurs with probability one.  Our preliminary results showing conditions for a stable embedding for linear systems of class $\mathcal{A}(1)$ were reported in~\cite{yap2010stabletakens}.

There has also been significant prior work related to embedding manifolds (or fractal sets), which has important implications for attractor embeddings.   Specifically, embedding results for manifolds were derived by Whitney \cite{whitney} and later expanded on by Sauer et al. \cite{embedology}. These results show that if a manifold has dimension $d$, then almost every smooth function mapping into $\reals^M$ with $M > 2d$ will be an embedding of the manifold. Baraniuk \& Wakin \cite{wakin_embedding} extended these results  to show that for manifolds with dimension $d$ embedded in $\reals^N$, random orthoprojections into $\reals^M$ provide a stable embedding of the manifold as long as $M$ scales linearly with $d$ and logarithmically with $N$  (depending also on various properties of the manifold, such as the maximum curvature).
Clarkson \cite{clarkson2008tighter} later improved on the required number of measurements $M$ by removing the dependence on $N$ and certain worst case properties of the manifold.  We note that these stable embedding results have been used to show that manifold learning and dimensionality estimation algorithms can be performed in the compressed space with nearly the same accuracy as they could be performed in the original space~\cite{hegde2007manlearn}.
The main distinction between these manifold embedding results and Takens' theorem is that these results acquire $M$ independent observations of each single point on the manifold, whereas Takens' result requires the repeated application of a single observation function to a system having its own internal time variations.  In essence, the delay coordinate map relies on the system dynamics to provide measurement diversity when the observations are restricted to a single fixed function $h$.

One of the principle benefits of a stable delay coordinate map would be resilience to noise and other imperfections. 
The effect of noise on the reconstruction of state space attractors has also been previously considered by several researchers apart from the notion of a stable embedding. In \cite{muldoon1998delay} the authors looked at a modified embedding theorem for systems corrupted by dynamical noise, considering specifically embeddings using multivariate time series system outputs and taking more measurements than is typically required for a delay-coordinate map.  In \cite{casdagli1991state}, the authors study the effects of observational noise via statistical methods, showing how the choice of delay-coordinates (i.e., the choice of observation function $h$ and sampling time $T_s$ with respect to the system dynamics) affects the ability to make predictions. In particular, they showed that poor reconstruction amplifies noise and increases estimation error.

In related work, there has also been considerable research on the choice of the optimal sampling interval $T_s$ for the construction of the delay coordinate map (typically for the study of chaotic dynamical systems). In particular, one of the more successful techniques is choosing $T_s$ to minimize the mutual information between any two time series samples separated by $T_s$~\cite{fraser1986independent}. The resulting reconstructed attractor usually makes the quantitative and qualitative study of the chaotic dynamics easier as the reconstructed trajectories tends to be unfolded to maximally fill the reconstruction space.  In contrast, our goal is to characterize conditions on the system and observation functions (including but not limited to $T_s$) such that the geometry of the attractor is faithfully represented in the reconstruction space.

\section{Stable Embeddings for Linear Dynamical Systems}

In this section we present our main technical results.  We first present a preliminary result in Section~\ref{sec:existence} that gives explicit sufficient conditions on the system and observation functions to guarantee that the delay coordinate map is a one-to-one map of the state space attractor.  This is akin to Takens' Embedding Theorem, and we present it here to highlight the specific differences that arise under our restrictions (linear systems and measurement functions) and when seeking explicit conditions on system and measurement pairs (as opposed to the conditions for generic observation functions in Takens' theorem).  We then present our main technical contribution in Section~\ref{sec:STE_Ad}, giving explicit conditions on the system and observation function for the delay coordinate map to  be a stable embedding of the attractor with specific guarantees on the conditioning number of the embedding.

\subsection{Takens' Embeddings}
\label{sec:existence}

The following theorem gives conditions on the system and the observation function such that the delay coordinate map $F$ is a one-to-one mapping.
This is analogous to Theorem~\ref{thm:takens} in the context of linear dynamical systems and linear observation functions.

\begin{thm}[Linear Takens' Embedding \cite{yap2010stabletakens}]
	\label{thm:existence_thm}
	Assume a linear dynamical system of class $\mathcal{A}(d)$ in $\reals^N$ that is in steady state. Choose $T_s > 0$ to be the sampling interval, $h \in \reals^N$ to be the observation function, and denote by $F$ the delay-coordinate map with $M$ delays as defined in \eqref{eq:DCM_linear_arb_h}. 
	Suppose that $M \ge 2d$, the $\mathcal{A}_{\Phi}$-eigenvalues $\{e^{\pm j\theta_i T_s}\}$ are distinct and strictly complex,\footnote{We say that a number $x$ is \emph{strictly complex} if $\Imo{x} \neq 0$.} and $v_i^H h \neq 0$ for all $i = 1, \cdots, d$.
	Then for all distinct pairs of points $x,y \in \mathcal{M}$, $F$ satisfies \eqref{eqn:ARIP} for some constants $C$ and $\delta < 1$.
\end{thm}

\emph{Proof: The proof of this theorem can be found in Appendix~\ref{sec:proof_main}.}

To explore the differences that arise in our specific setting of linear systems and linear observation functions, we compare the conditions of this theorem with that of Takens' theorem.  First, we notice that the conditions on the measurement operation are very similar.  Theorem~\ref{thm:existence_thm} requires $M \ge 2d$, which is similar to Takens' $M>2d$ and likely only different because of the specific structure of our attractors.  There is also a close correspondence with the other condition on the measurement function $v_i^H h \neq 0$.  This requirement is an explicit condition on the relationship between the system and observation function ensuring that the observation function can capture some information from every dimension of the attractor.  We note that (Lebesgue) almost-every $h \in \reals^N$ will satisfy this condition, and so we find that this is just a more explicit version of Takens' result that ``almost-every'' $h$ ensures an embedding.

Next, we compare our conditions on the system with those imposed by Takens' theorem.  Theorem~\ref{thm:existence_thm} requires that the $\mathcal{A}_{\Phi}$-eigenvalues are distinct and strictly complex, which is equivalent to having $e^{j\theta_p T_s} \neq e^{\pm j \theta_q T_s}$ (distinct) and $e^{j\theta_p T_s} \neq \pm 1$ (strictly complex) for all $p \neq q$ and $p,q = 1, \cdots, d$.  
While this requirement implies\footnote{This implication can be shown by contradiction. Pick any $1 \le k \le 2d$ and suppose that $\mathcal{M}$ has at least a periodic orbit of $\Psi$ with period $kT_s$. This would be equivalent to saying that $e^{j \theta_p kT_s} = \left(e^{j \theta_p T_s}\right)^{k} = 1$ for all $p$, meaning that for each $p$ from 1 to $d$ the quantity $e^{\pm j \theta_p T_s}$ is uniquely one of the $k$ roots of unity. However this is impossible as there are $2d$ distinct and strictly complex values of $\{e^{\pm j\theta_p T_s}\}$ and there are only $k \le 2d$ roots of unity (including $\pm 1$ which are not allowed), and hence we have a contradiction.} that $\mathcal{M}$ does not have periodic orbits of period $kT_s$ for $k = 1, \cdots, 2d$  (thus satisfying Takens' condition), 
our condition is actually more stringent than this restriction on periodic orbits (likely due to our restricted class of linear observation functions).
We note that since $\{\theta_i\}_{i=1}^{d}$ are distinct by definition, this condition is dependent on the choice of sampling interval $T_s$. One can verify that choosing $T_s < \frac{\pi}{\max\{ \theta_i \}}$ is sufficient (but not necessary) to meet the condition of the theorem.

\subsection{Stable Takens' Embeddings}
\label{sec:STE_Ad}

Before presenting our main result giving conditions for a stable embedding of a dynamical system in a delay coordinate map, it will be useful to define and understand the following quantities that characterize how well-behaved the system and measurement process are both individually and jointly.  First, we define 
$\kappa_1 = \min_{i \in \{1, \dots, d\}} \left\{\frac{|v_i^H h|}{\|h\|_2} \right\}$ and $\kappa_2 = \max_{i \in \{1,\dots, d\}} \left\{\frac{|v_i^H h|}{\|h\|_2} \right\}$ 
characterizing the minimum and maximum projection of the (normalized) observation function on the $\mathcal{A}$-eigenvectors. Roughly speaking, these quantities are an indication of the disparity between the dimensions of the system attractor that are best and worst matched to the observation function.  One would expect that a measurement system is most efficient when it observes all parts of the attractor equally such that $\kappa_1 \approx \kappa_2$.  Second, we define $A_1, A_2$ as the smallest and largest eigenvalues of $V^H V$, respectively.  As we discussed at the end of Section \ref{sec:linear_sys}, these quantities describe how well the system attractor fills the dimensions of the state space that it occupies (i.e., when $A_2\gg A_1$ the attractor is very elongated in the state space).  Again, we would expect that a system will be most amenable to observation when it fills the space such that $A_1\approx A_2$. 

Finally, we define
$\nu := \underset{p \neq q}{\max} \left\{ {\left|\sin(\theta_p T_s)\right|}^{-1}, 
{\left| \sin\left( \frac{(\theta_p - \theta_q)T_s}{2} \right) \right|}^{-1},
{\left| \sin\left( \frac{(\theta_p + \theta_q)T_s}{2} \right) \right|}^{-1} \right\}$, 
which will also bound the constants associated with the stable embedding.  Notice that the first term is large if $\theta_p T_s$ is small for some $p$ (or that $\theta_p T_s \approx k\pi$ for some integer $k$), meaning that the system state proceeds  in the span of $\Reo{v_p}$ and $\Imo{v_p}$ at a slow pace, thus not producing much diversity in consecutive measurements of the system along these dimensions.   
The second term is large if $\theta_p T_s - \theta_q T_s$ is small (or near $k\pi$) for some $p \neq q$ and $p,q = 1, \cdots, d$, implying that the system state is proceeding  in the subspaces spanned by $\Reo{v_p}, \Imo{v_p}$ and $\Reo{v_q}, \Imo{v_q}$ at almost the same rate. %
This condition would be unfavorable because the system will take an extremely long time to display enough diversity to determine that it is actually traveling on two separate subspaces instead of one.  The third term is similar to the second term if we write $\theta_p T_s + \theta_q T_s = \theta_p T_s - (-\theta_q T_s)$. Thus if $\theta_p T_s \sim -\theta_q T_s$, then the system is again proceeding on two subspaces at almost the same rate (although the system is proceeding in one of the subspaces in the ``opposite'' direction).

Armed with these definitions, we now present our main result giving deterministic, explicit and  non-asymptotic guarantees on the conditioning of the delay coordinate map.
\begin{thm}[Stable Linear Takens' Embedding]
	\label{thm:stability_thm}
	Assume a linear dynamical system of class $\mathcal{A}(d)$ in $\reals^N$ that is in steady state. Choose $T_s > 0$ to be the sampling interval, $h \in \reals^N$ to be the observation function  such that $\|h\|_2^2 = \frac{2d}{M}$, and denote by $F$ the delay-coordinate map with $M$ delays as defined in \eqref{eq:DCM_linear_arb_h}. 
	Suppose that $M~>~\left((2d~-~1)~\frac{A_2 \kappa_2^2}{A_1 \kappa_1^2}~\nu\right)$,
	the $\mathcal{A}_{\Phi}$-eigenvalues $\{e^{\pm j\theta_i T_s}\}$ are distinct and strictly complex, and
	$v_i^H h \neq 0$ for all $i = 1, \cdots, d$.
	Then  for all distinct pairs of points $x,y \in \mathcal{M}$, $F$ satisfies \eqref{eqn:ARIP} with constants $C~:=~d~\left( \frac{\kappa_1^2}{A_2} + \frac{\kappa_2^2}{A_1} \right)$ and $\delta := \delta_0 + \delta_1(M)$, where:
	\begin{eqnarray}
		\delta_0 := \frac{A_2 \kappa_2^2 - A_1 \kappa_1^2 }{A_2 \kappa_2^2 + A_1 \kappa_1^2}, \;\;\;\;\; 
		\delta_1(M) := \frac{(2d-1) \nu}{M} \left(\frac{2 A_2 \kappa_2^2}{A_2 \kappa_2^2 + A_1 \kappa_1^2}\right). \label{eqn:delta1}
	\end{eqnarray}
\end{thm}

\emph{Proof: The proof of this theorem can be found in Appendix \ref{sec:proof_main}.}

We first note that the sufficient conditions of this theorem are the same as those in Theorem~\ref{thm:existence_thm}, except that the required number of measurements is larger to ensure specific guarantees on the conditioning number $\delta$ (i.e. $\delta < 1$).  Also, note that this theorem requires an observation function with a particular norm $\|h\|_2^2 = \frac{2d}{M}$.   This normalization is to remove from $C$ any dependence on the number of measurements $M$ and the dimension of the attractor $2d$ (since $\kappa_1^2$ and $\kappa_2^2$ both scale inversely with $d$).  The normalization plays no other significant role in the proof (and therefore could be eliminated without losing generality, but at the expense of clarity).

To understand the implications of Theorem~\ref{thm:stability_thm}, we examine the behavior of the conditioning number $\delta$ as it is the main quantity of interest.  In the theorem statement, $\delta$ is a sum of $\delta_0$ (which does not depend on $M$) and $\delta_1(M)$ which is positive for all $M$ and for which $\lim_{M\to \infty} \delta_1(M)=0$.  Thus, we see that by taking more observations one could drive the conditioning guarantee for the mapping to $\delta=\delta_0$, \emph{but not below}.  In other words, some system and measurement pairs will have a plateau preventing the conditioning guarantee for the delay coordinate map from improving beyond a fundamental limit.  This is in contrast with CS results where the conditioning can be continually improved by taking more measurements.  Indeed, in order to get arbitrarily good conditioning we would need $\delta_0=0$, which happens if and only if
	$A_2 \kappa_2^2 - A_1 \kappa_1^2 = 0 \; \Leftrightarrow \; \frac{A_2}{A_1} = \frac{\kappa_1^2}{\kappa_2^2}=1$.
Recall that $A_1 = A_2$ implies that the attractor $\mathcal{M}$ maximally fills the subspace spanned by $V$ 
and $\kappa_1 = \kappa_2$ means that the observation function $h$ projects equally onto the $\mathcal{A}$-eigenvectors. Thus even with an infinite number of measurements, the delay coordinate map can only be guaranteed to be an exact isometry ($\delta=0$) when the system and observation function maximally fill and measure the subspace containing the attractor.

The quantity $\delta_1(M)$ can be used to determine the number of measurements necessary to ensure that the conditioning number $\delta$ is within $\epsilon$ of the optimal value $\delta_0$.  To find the required number of measurements to meet this target $\widehat{M}(\epsilon)$, we set $\delta_1(M)=\epsilon$ and solve \eqref{eqn:delta1} for $M$ to get
\begin{eqnarray}
	\widehat{M}(\epsilon) = \frac{(2d-1) \nu}{\epsilon} \left(\frac{2 A_2 \kappa_2^2}{A_2 \kappa_2^2 + A_1 \kappa_1^2}\right).
	\label{eq:def_widehatM}
\end{eqnarray}
By multiplying the numerator and denominator by $\frac{1}{A_2 \kappa_2^2}$ and noting that $0 < \frac{A_1 \kappa_1^2}{A_2 \kappa_2^2}  \le 1$, we can deduce that $\frac{(2d-1) \nu}{\epsilon} \le \widehat{M}(\epsilon) < \frac{2(2d-1) \nu}{\epsilon}$.  One immediate application of this fact is that we can calculate the number of measurements necessary to guarantee a stable embedding for the delay coordinate map with a specified conditioning $\delta \in (\delta_0, \; 1)$, which is made precise in the following corollary.

\begin{cor}
Suppose we have a linear system of class $\mathcal{A}(d)$, observation function $h$ and sampling time $T_s$ such that the conditions of Theorem \ref{thm:stability_thm} are satisfied.  Choose any 
$0<\epsilon<\left(1-\delta_0\right)$. 
If the delay coordinate map $F$ defined in \eqref{eq:DCM_linear_arb_h} has a number of delays $M$ chosen to satisfy $M \ge \frac{2(2d - 1)\nu}{\epsilon}$, then $F$ is a stable embedding of $\mathcal{M}$ with conditioning $\delta \leq \delta_0 + \epsilon$.
\end{cor}

The proof of this corollary is not shown, but follows immediately from Theorem \ref{thm:stability_thm}.  While the linear scaling with $d$ seen in this result is in line with state-of-the-art CS results, we see that in contrast to typical CS results    $\widehat{M}(\epsilon)$ does not depend on the ambient dimension $N$.  Also note that $\widehat{M}(\epsilon)$ depends strongly on the $\mathcal{A}$-eigenvalues via the quantity $\nu$. In contrast, the interactions of the $\mathcal{A}$-eigenvectors and the observation function $h$ determine the lower bound on the conditioning $\delta$, as evidenced by the roles played by the quantities $A_1,A_2$ and $\kappa_1, \kappa_2$ in the formula for $\delta_0$.

\section{Simulation experiments}
\label{sec:sims}

While the main result in Theorem~\ref{thm:stability_thm} is encouraging, it remains to be shown that $(i)$ the  theoretical quantities actually reflect the salient embedding characteristics seen in system and measurement combinations, and $(ii)$ having a stable embedding actually improves our ability to infer information about a hidden attractor. For example, it is important to know if the fundamental limits on the embedding quality $\delta(M)$ are artifacts of our proof technique or are empirically observed.  If these limits on the embedding quality are actually present, it is also important to know if the related bounds are tight, both in their asymptotic values and in terms of their convergence speed as $M$ increases.  Finally, for a stable embedding to be a valuable goal, we need to demonstrate that achieving this goal results in improved performance in specific tasks performed in the reconstruction space.  This section will use a series of simple simulations to explore these aspects of our theoretical results.

As a general approach, each simulation in Sections~\ref{sec:bounds} and~\ref{sec:convsp} below involve creating an observation function $h$ and a test system of dimension $N=50$ in class $\mathcal{A}(d)$ (defined by $\mathcal{A}$-eigenvalues and $\mathcal{A}$-eigenvectors) so that the conditions of Theorem~\ref{thm:stability_thm} are satisfied.  
We choose the arbitrary initial point $x_0$  defining the attractor such that $\alpha_0 = [1, \; \cdots,\; 1]^T$ and $x_0 = V \alpha_0$, and we assume a sample time of $T_s=1$. For a single trial, we generate a random pair of points on the attractor $x$ and $y$ by choosing uniform random numbers $t_x, t_y$ from $(0, 10000)$ and assigning $x = V e^{\Lambda t_x} \alpha_0$ and $y = V e^{\Lambda t_y} \alpha_0$.  In other words, we start the system from the (arbitrary) initial condition and stop it after a random amount of time to get a single point on the attractor.  We then vary $M$ from 1 to 200, and run 1000 trials for each $M$ (renormalizing $h$ for each $M$ as per Theorem~\ref{thm:stability_thm}).  For each trial we calculate the quality of the conditioning  $Q(x,y) = \frac{\|F(x) - F(y)\|_2^2}{\|x-y\|_2^2}$, and for each $M$ record the largest and smallest value of $Q(x,y)$ (denoted $\max\{Q\}$ and $\min\{Q\}$, respectively) as a way to quantify how the conditioning changes with the number of measurements.  In the subsequent plots the dotted lines represent $C(1\pm\delta_0)$, showing the theoretical asymptotic bounds on the conditioning quality $Q(x,y)$, and the dashed lines are the theoretical bounds on the conditioning $C(1 \pm \delta(M))$ given by Theorem~\ref{thm:stability_thm}.

\subsection{Bounds on the embedding quality}
\label{sec:bounds}

\begin{figure*}
	\hfil
	\begin{minipage}[t]{0.32\linewidth}
	\centerline{\epsfysize = 44mm \epsffile{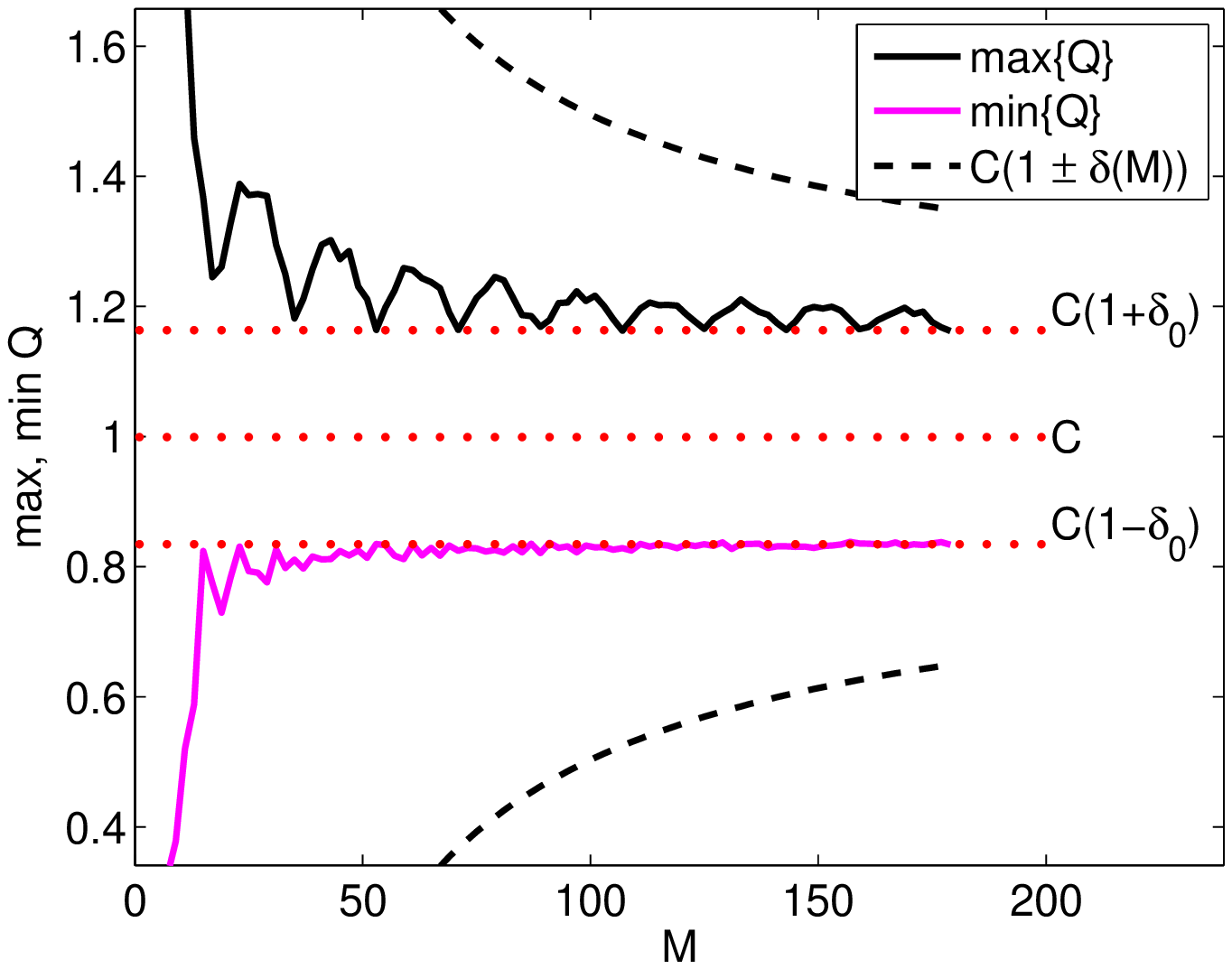}}
	\vspace{-1mm}
	\centerline{\small$\quad$~(a)}
	\end{minipage}
	%
	\hfil
	\begin{minipage}[t]{0.32\linewidth}
	\centerline{\epsfysize = 44mm \epsffile{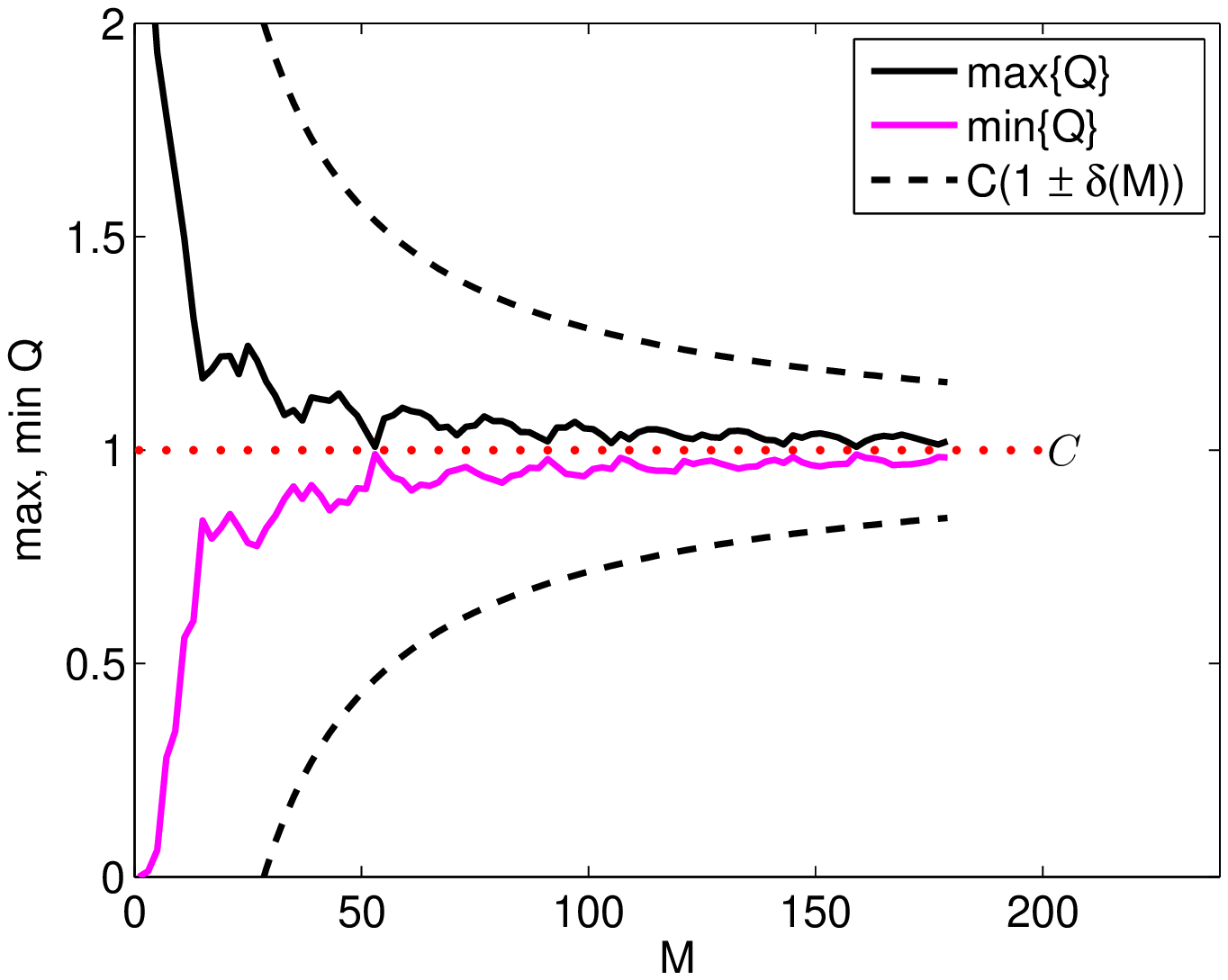}}
	\vspace{-1mm}
	\centerline{\small$\quad$~(b)}
	\end{minipage}
	\hfil
	\begin{minipage}[t]{0.32\linewidth}
	\centerline{\epsfysize = 44mm \epsffile{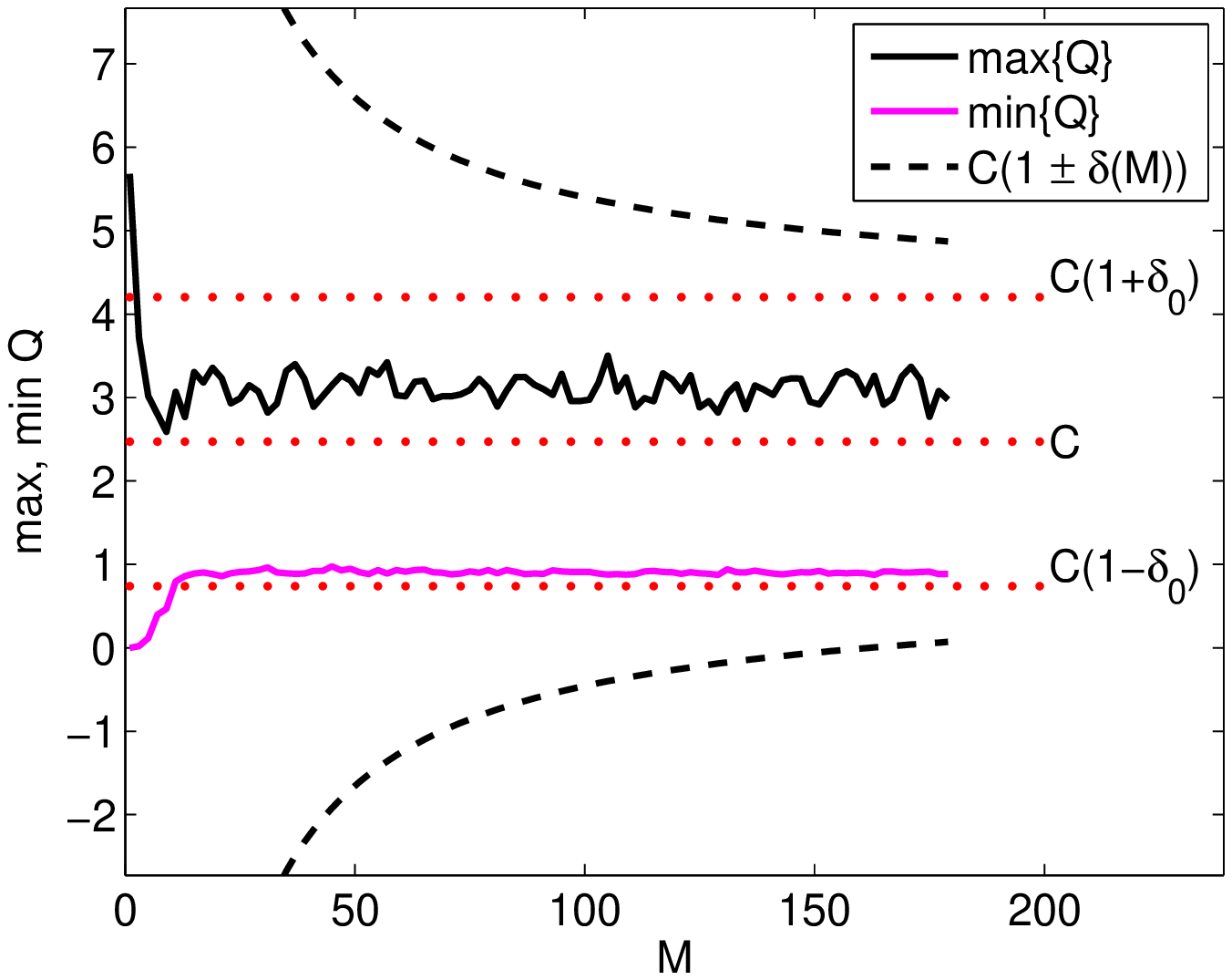}}
	\vspace{-1mm}
	\centerline{\small$\quad$~(c)}
	\end{minipage}
	\vspace{-1mm}
	\caption{\sl\small 
	Simulations exploring the asymptotic bounds on the conditioning of the delay coordinate map.  Plotted are the largest and smallest value of $Q(x,y)$ (depicted by $\max\{Q\}$ and $\min\{Q\}$ respectively) attained by the 1000 pairs of $x,y$ for each $M$. The dotted (red) lines represent the values of $C(1\pm\delta_0)$ and $C$, and the dashed (black) lines are the theoretical values of $C(1 \pm \delta(M))$. 
	(a) In this simulation, $A_1 = A_2$ but $\kappa_1 \neq \kappa_2$, thus a plateau  on the conditioning is seen.
	(b) In this simulation, $A_1 = A_2$ and $\kappa_1 = \kappa_2$. As expected, the conditioning number asymptotically reaches 0 as $M$ grows.
	(c) In this simulation, $A_1 \neq A_2$ and $\kappa_1 \neq \kappa_2$ and the predicted asymptotic values of the conditioning are not tight.}
	\label{fig:edu_STE}\label{fig:edu2_STE}
	\vspace{-8mm}
\end{figure*}

One of the fundamental characteristics of Theorem~\ref{thm:stability_thm} is that in general, the bound on the embedding quality $\delta (M)$ approaches $\delta_0\neq0$ as $M$ increases rather than approaching zero as is typical in CS results.  The first question to ask is whether pairs of systems and observation functions can actually display such a plateau as predicted, or whether the conditioning instead continually improves with more measurements.  To demonstrate this effect, we generate a simulation as described above with $d = 3$, choosing the $\mathcal{A}$-eigenvalues $\{\theta_i\}_{i=1}^{d}$ uniformly at random from $(0,\pi)$, and taking care to ensure that 
the resulting $\mathcal{A}_{\Phi}$-eigenvalues are distinct and strictly complex to satisfy the conditions of Theorem~\ref{thm:stability_thm}. We then create the $\mathcal{A}$-eigenvectors by letting
	$v_i = \frac{1}{\sqrt{2}} (e_{2i-1} + j e_{2i})$,
where $\{e_i\}$ are the canonical basis vectors in $\reals^{N}$.   This choice of $\mathcal{A}$-eigenvectors ensures that  $A_1 = A_2$.  To generate a generic observation function $h$, we first create a vector $c \in \reals^N$ such that $c = \sum_{i = 1}^{d}((1 + w_{2i-1}) \Reo{v_i} +  (1 + w_{2i}) \Imo{v_i})$, where the $\{w_i\}$ are i.i.d. Gaussian random variables of zero mean and variance $0.1$.
Thus $c$ is a (random) linear combination of the vectors that form the subspace of the attractor. 
For each $M$ we let $h = h(M) = \sqrt{\frac{2d}{M}}\frac{c}{\|c\|_2}$ so that $\|h\|_2^2 = \frac{2d}{M}$ to meet the conditions of Theorem~\ref{thm:stability_thm}.   Note that the small variance of $\{w_i\}$ produces  $\{{|v_i^H h|^2}/{\|h\|_2^2}\}$ centered tightly around 1, making $\delta_0$ small (due to $A_1=A_2$ and $\kappa_1$, $\kappa_2$ both close to 1).\footnote{The random variables $\{w_i\}$ are used to ensure that $\kappa_1$, $\kappa_2$ are close to, but not exactly equal to 1.  The case where $\kappa_1=\kappa_2=1$ is considered in the simulation in Figure \ref{fig:edu_STE}(b).}  The specific parameters in this simulation are shown in Table~\ref{tab:ex_d3_Ae_Kne}. 

\begin{table*}[th]
	\small
	\begin{center}
 \begin{tabular}{|c|c|c|c|c|c|c|}
 \hline
 Index $i$ & 1 & 2 & 3 & 4 & 5 & 6 \\
 \hline
 $\theta_i$ (rad) & 2.3129 & 0.1765 & 1.4861 & --- & --- & --- \\
\hline
${|v_i^H h|^2}/{\|h\|_2^2}$ & 0.8346 & 1.1637 & 1.0017 & --- & --- & --- \\
\hline
$\lambda_i(V^H V)$ & 1 & 1 & 1 & 1 & 1 & 1 \\
\hline
\end{tabular}
\end{center}
\vspace{-5mm}
\caption{\small\sl{Parameters for the simulation shown in Figure~\ref{fig:edu_STE}(a). In this case the relevant quantities are $A_1 = A_2 = 1$, $\kappa_1 = 0.8346$, $\kappa_2 = 1.1637$, $\nu = 5.6954$ and $\delta_0 = 0.1647$.}}
\label{tab:ex_d3_Ae_Kne}
\vspace{-12mm}
\end{table*}

The results for this simulation are shown in Figure~\ref{fig:edu_STE}(a).  We see from the behavior of $\max\{Q\}$ and $\min\{Q\}$ that the embedding does indeed reach a fundamental limit where the conditioning does not improve with more measurements.  Furthermore, we see in this case that this plateau is correctly captured by the value $C(1\pm\delta_0)$ as described in Theorem~\ref{thm:stability_thm}.  Additionally, the bounds $C(1 \pm \delta(M))$ do contain $\max\{Q\}$ and $\min\{Q\}$ as expected from the theorem, and the characteristic shape of these curves seems to qualitatively reflect the empirically observed convergence of the conditioning number.

As confirmation, we also verify the implication of Theorem~\ref{thm:stability_thm} that system and measurement combinations can be constructed where the conditioning can be made arbitrarily good with more measurements (akin to the more typical CS results).  To show this, we create another system with the same $\mathcal{A}$-eigenvalues and $\mathcal{A}$-eigenvectors as in the previous simulation, with the latter implying that $A_1 = A_2$. 
For the observation function, we first define $c = V [1,\; \cdots,\; 1]^T$, and for each $M$ we let $h = h(M) = \sqrt{\frac{2d}{M}} \frac{c}{\|c\|_2}$ as before. One can verify this choice results in $|v_i^H h|/\|h\|_2 = 1$ for all $i$, and thus $\kappa_1 = \kappa_2$. The parameters of this experiment are summarized in Table \ref{tab:ex_d3_Ae_Ke}.

\begin{table*}[th]
	\small
	\begin{center}
 \begin{tabular}{|c|c|c|c|c|c|c|}
 \hline
 Index $i$ & 1 & 2 & 3 & 4 & 5 & 6 \\
 \hline
 $\theta_i$ (rad) & 2.3129 & 0.1765 & 1.4861 & --- & --- & --- \\
\hline
${|v_i^H h|^2}/{\|h\|_2^2}$ & 1 & 1 & 1 & --- & --- & --- \\
\hline
$\lambda_i(V^H V)$ & 1 & 1 & 1 & 1 & 1 & 1 \\
\hline
\end{tabular}
\end{center}
\vspace{-5mm}
\caption{\small\sl{Parameters for the simulation shown in Figure \ref{fig:edu_STE}(b). The experiment was chosen such that $A_1 = A_2 = 1$ and $\kappa_1 = \kappa_2 = 1$, so that $\delta_0 = 0$. As the $\mathcal{A}$-eigenvalues are the same as in the previous experiment, $\nu$ remains at $5.6954$.}}
\label{tab:ex_d3_Ae_Ke}
\vspace{-12mm}
\end{table*}

With this choice of parameters such that $A_1 = A_2$ and $\kappa_1 = \kappa_2$,  Theorem~\ref{thm:stability_thm} indicates that $\delta_0=0$ so that ${\lim_{M\to\infty} \delta(M)= 0}$.  Figure~\ref{fig:edu_STE}(b) shows the results of running the simulation in the same manner as before.  The values of $\max\{Q\}$ and $\min\{Q\}$ clearly converge to $C$ as expected, showing that in this case the conditioning of the embedding can indeed be made arbitrarily good by taking more measurements.

Although Theorem~\ref{thm:stability_thm} indicates that a finite limit on the conditioning number is always reached when either $A_1 \neq A_2$ or $\kappa_1 \neq \kappa_2$, this bound is not always tight and the predicted plateau level of $C(1 \pm \delta_0)$ may be conservative. 
To show this, we construct a similar simulation as above, now setting the $\mathcal{A}$-eigenvectors to be
	$v_i = \frac{1}{\sqrt{\|a_i\|_2^2 + \|b_i\|_2^2}} (a_{i} + j b_{i})$,
where $\{a_i,b_i\}$ are randomly constructed vectors in $\reals^N$ whose entries are i.i.d. zero-mean Gaussian random variables with a variance of $1$.
We keep the $\mathcal{A}$-eigenvalues the same and generate $h$ in the same manner as the first simulation shown in Figure~\ref{fig:edu_STE}(a).  The specific parameters for this simulation are shown in Table \ref{tab:ex_d3_Ane_Kne}, where we see that indeed $A_1 \neq A_2$ and $\kappa_1 \neq \kappa_2$.
Figure \ref{fig:edu2_STE}(c) shows the results of running the simulation in the same manner as before.  We see that although a limit on the conditioning number is reached as predicted by Theorem~\ref{thm:stability_thm}, the predicted plateau level of $C(1\pm\delta_0)$ is not tight and the conditioning can be better than that predicted by $\delta_0$.

\begin{table*}[th]
	\small
	\begin{center}
 \begin{tabular}{|c|c|c|c|c|c|c|}
 \hline
 Index $i$ & 1 & 2 & 3 & 4 & 5 & 6 \\
 \hline
 $\theta_i$ (rad) & 2.3129 & 0.1765 & 1.4861 & --- & --- & --- \\
\hline
${|v_i^H h|^2}/{\|h\|_2^2}$ & 1.8138  &  1.2064  &  1.1318 & --- & --- & --- \\
\hline
$\lambda_i(V^H V)$ &  1.5316 &   1.3058 &   1.1294  &  0.8372  &  0.7644  &  0.4315 \\
\hline
\end{tabular}
\end{center}
\vspace{-5mm}
\caption{\small\sl{Parameters for the simulation shown in Figure \ref{fig:edu2_STE}(c). We see that $A_1 = 0.4315$, $A_2 = 1.5316$, $\kappa_1 = 1.1318$ and $\kappa_2 = 1.8138$. Since the $\mathcal{A}$-eigenvalues are the same as in the first simulation shown in Figure~\ref{fig:edu_STE}(a), $\nu$ remains the same at $5.6954$. We also calculate  $\delta_0 = 0.7010$.}}
\label{tab:ex_d3_Ane_Kne}
\vspace{-12mm}
\end{table*}

\subsection{Convergence Speed}
\label{sec:convsp}

In the simulations of the previous section we concentrated on the conditioning limits predicted by Theorem~\ref{thm:stability_thm}, ignoring
issues of the speed of convergence to those limits. 
Examining the formula for $\delta_1(M)$ in Theorem~\ref{thm:stability_thm}, we see that the $\mathcal{A}$-eigenvalues (via the parameter $\nu$) affect the convergence speed of $\delta(M)$ to its asymptotic value of $\delta_0$.  In particular, the convergence speed scales with  $1/\nu$, which is also demonstrated in \eqref{eq:def_widehatM} where the number of measurements $\widehat{M}(\epsilon)$ necessary to get the conditioning $\delta$ within $\epsilon$ of the best possible value  $(\delta_0)$ is proportional to $\nu$.

\begin{figure*}
	\hfil
	\begin{minipage}[t]{0.4\linewidth}
	\centerline{\epsfysize = 50mm \epsffile{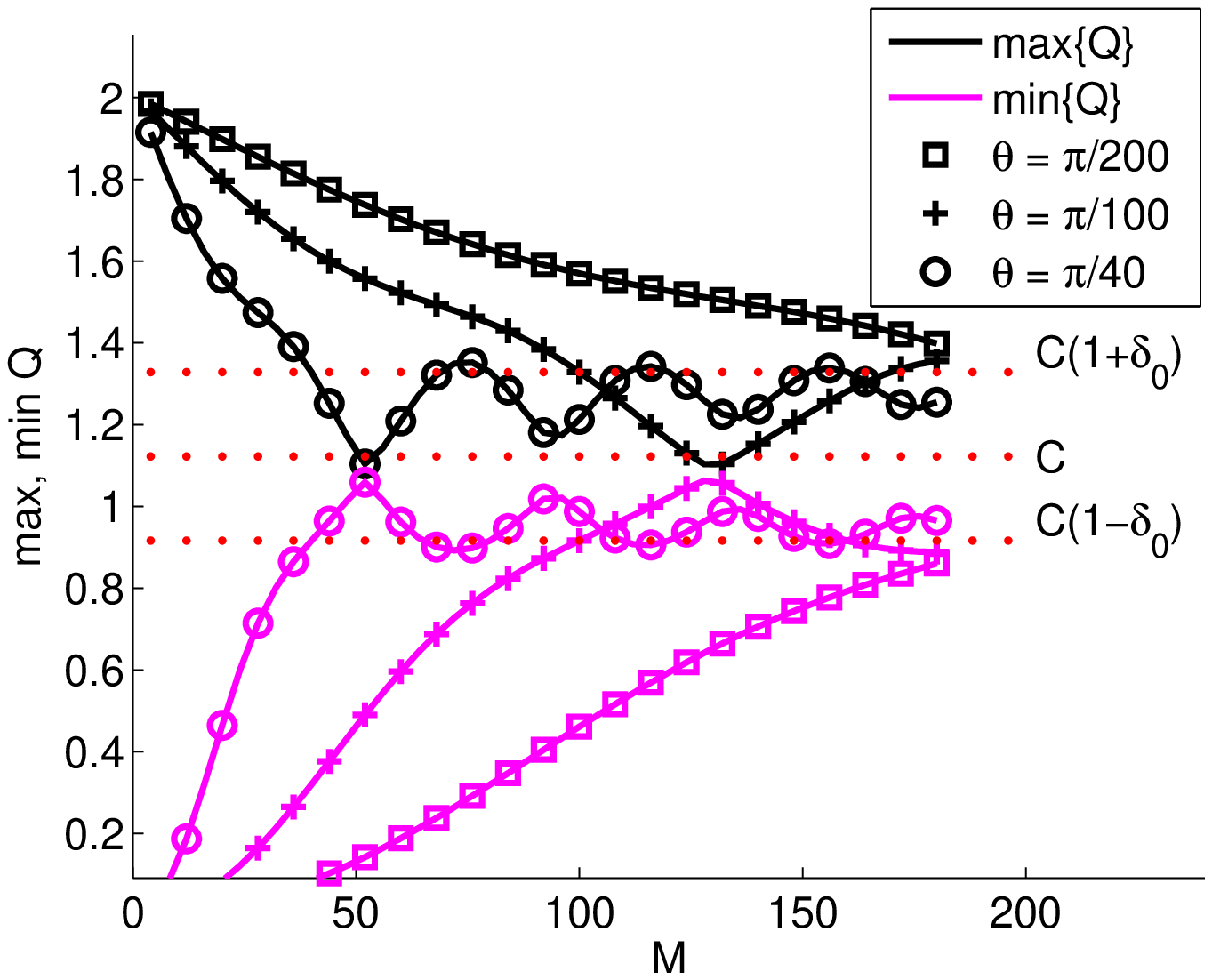}}
	\vspace{-1mm}
	\centerline{\small$\quad$~(a)}
	\end{minipage}
	\hfil
	\begin{minipage}[t]{0.4\linewidth}
	\centerline{\epsfysize = 50mm \epsffile{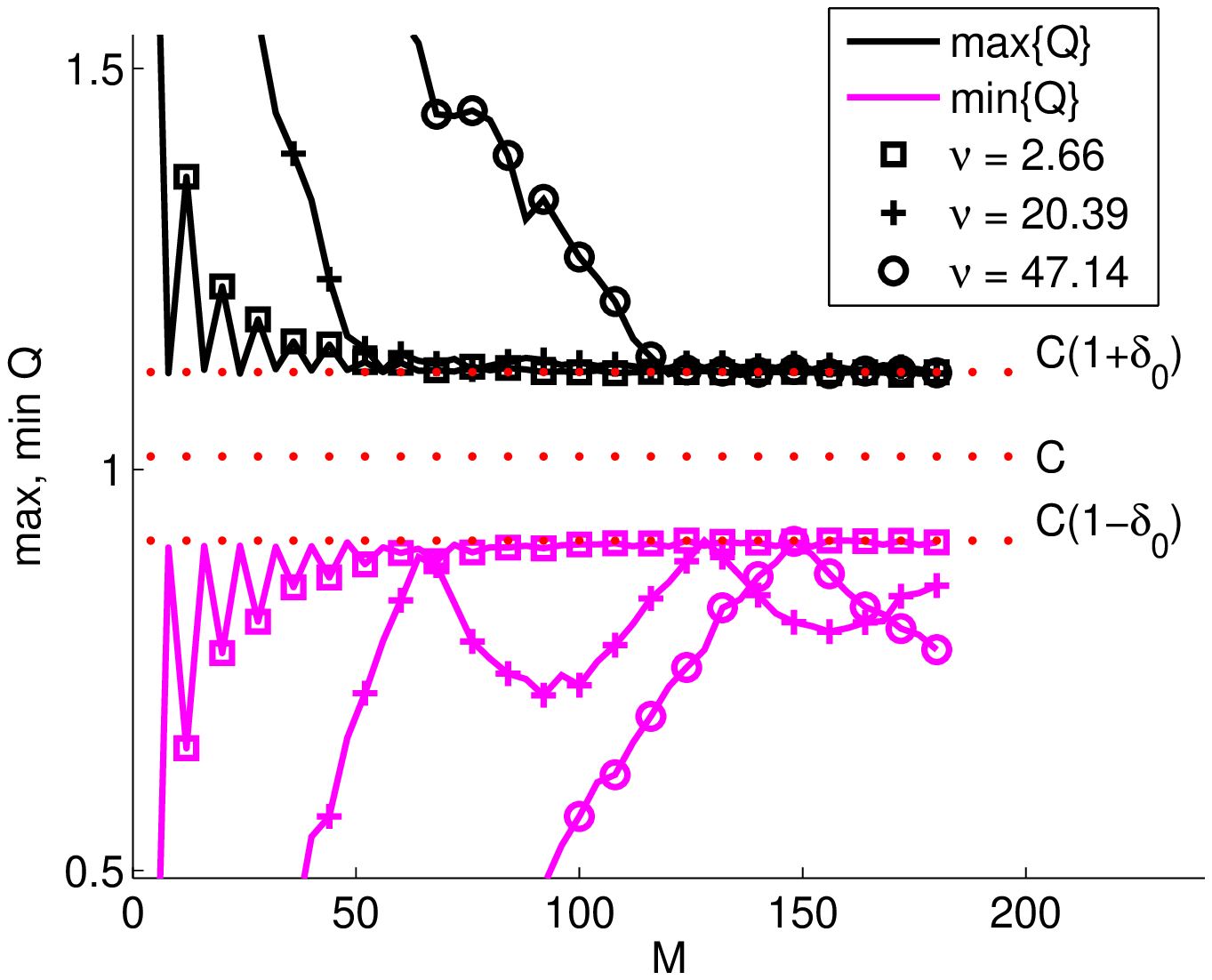}}
	\vspace{-1mm}
	\centerline{\small$\quad$~(b)}
	\end{minipage}
	\vspace{-1mm}
	\caption{\sl\small 
	Examining the effect of the $\mathcal{A}$-eigenvalues on the convergence speed of the conditioning. (a) In this simulation, $d = 1$ and we test $\theta=\frac{\pi}{200}, \frac{\pi}{100}$ and $\frac{\pi}{40}$. 
	As expected, the closer $\theta$ is to $\pi/2$, the faster the rate of convergence of $\delta(M)$ to $\delta_0$.
	(b) In this simulation, $d = 3$ and we vary between 3 sets of $\mathcal{A}$-eigenvalues with different values of $\nu$. 
	As expected, the set of eigenvalues that gives the smallest $\nu$ provides the fastest rate of convergence of $\delta(M)$ to $\delta_0$ and vice versa.}
	\label{fig:d1_3theta_STE}
	\vspace{-8mm}
\end{figure*}

For ease of analysis, we first consider the case where $d=1$, meaning that $\nu=|\sin(\theta)|^{-1}$ (since $T_s = 1$), where $\pm j\theta$ are the sole $\mathcal{A}$-eigenvalues.  
In this case, $|\sin(\theta)|^{-1} \ge 1$ with the minimum attained when $\theta = \frac{\pi}{2} +k\pi$ for any integer $k$.  The closer $\theta$ is to $\frac{\pi}{2} +k\pi$, the faster the convergence of $\delta(M)$ to $\delta_0$.   
This is illustrated by the following simulation where the $\mathcal{A}$-eigenvectors are chosen such that $A_1 = A_2$, and the observation function is chosen randomly as in the experiment shown in Figure~\ref{fig:edu2_STE}(a) (except with $d=1$).  Figure~\ref{fig:d1_3theta_STE}(a) plots $\max\{Q\}$ and $\min\{Q\}$ for $\theta=\frac{\pi}{200}, \frac{\pi}{100}$ and $\frac{\pi}{40}$, showing that Theorem~\ref{thm:stability_thm} correctly captures that the convergence speed to the asymptotic value of $C(1 \pm \delta_0)$ varies inversely with the value of $\theta$.

When $d > 1$, the joint relationship of the $\mathcal{A}$-eigenvalues (not just their individual values) determines $\nu$, and subsequently the convergence speed.  One can see intuitively in the definition of $\nu$  that $\mathcal{A}$-eigenvalues which are maximally spread out should produce favorable convergence speeds. 
To illustrate this, we generate a simulated system with $d=3$, choosing the $\mathcal{A}$-eigenvectors such that $A_1 = A_2$, and generating an observation function $h$ randomly (as in the experiment in Figure \ref{fig:edu_STE}(a)).  We also choose three sets of $\mathcal{A}$-eigenvalues: two uniformly random sets, and one set that are slight perturbations of equally spaced points around the unit circle according to $\theta_p = \frac{p \pi}{d+1}$ (the choices of $\theta_p$ and their respective $\nu$ are given in Table~\ref{tab:d3_3nu_STE}).\footnote{The slight perturbation is used for plotting convenience so all three curves converge to the same asymptotic value.  If exactly equally spaced eigenvalues are used, the attractor is sampled uniformly and the convergent value will be inside $C(1 \pm \delta_0)$, making comparative plots difficult.}  Figure \ref{fig:d1_3theta_STE}(b) shows the results of the simulation, with the $\max\{Q\}$ and $\min\{Q\}$ curves showing clearly that  $\nu$ indeed controls the speed of convergence of $\delta(M)$ as predicted.

\begin{table*}[th]
	\small
	\begin{center}
 \begin{tabular}{|c|c|c|c|c|}
 \hline
  & $\theta_1$ & $\theta_2$ & $\theta_3$ & $\nu$\\
 \hline
 Set 1 (nearly equal spacing) & 0.7836 & 1.5864 & 2.3566 & 2.6619 \\
\hline
 Set 2 (random) & 0.0491 & 1.5737 & 2.3490 & 20.3851 \\
\hline
 Set 3 (random) & 0.0212 & 1.5684 & 2.3549 & 47.1388 \\
\hline
\end{tabular}
\end{center}
\vspace{-5mm}
\caption{\small\sl{Choice of $\{\theta_i\}$ (in radians) for the experiment in Figure \ref{fig:d1_3theta_STE}(b) and their respective $\nu$ value.}}
\label{tab:d3_3nu_STE}
\vspace{-12mm}
\end{table*}

\begin{figure*}
	\hfil
	\begin{minipage}[t]{0.4\linewidth}
	\centerline{\epsfysize = 50mm \epsffile{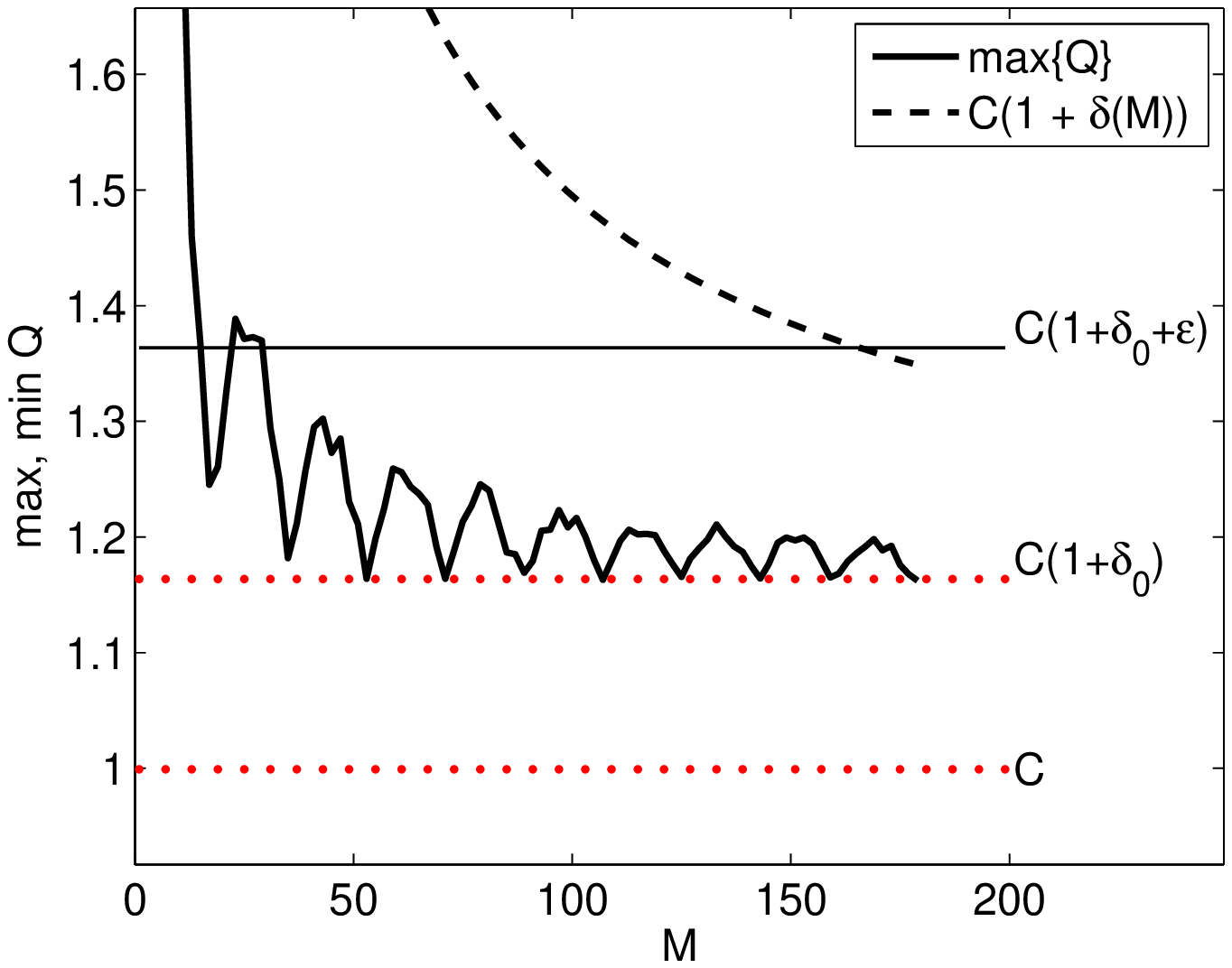}}
	\vspace{-1mm}
	\centerline{\small$\quad$~(a)}
	\end{minipage}
	\hfil
	\begin{minipage}[t]{0.4\linewidth}
	\centerline{\epsfysize = 50mm \epsffile{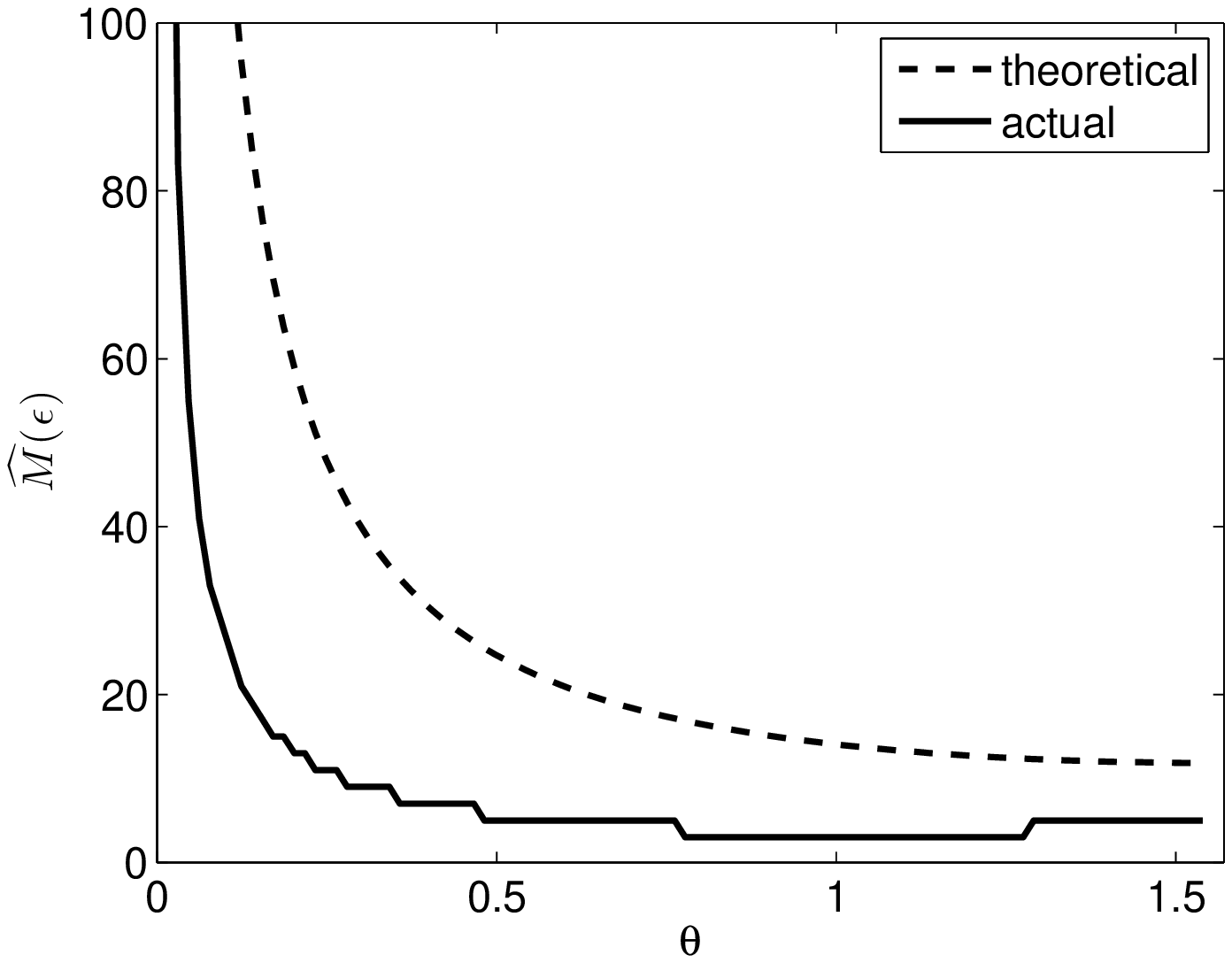}}
	\vspace{-1mm}
	\centerline{\small$\quad$~(b)}
	\end{minipage}
	\vspace{-1mm}
	\caption{\sl\small  
	Examining the predicted number of measurements necessary to reach a specified conditioning level.  (a) Plotted is the upper-half of Figure \ref{fig:edu_STE}(a), also indicating $C(1+\delta_0 + \epsilon)$ with $\epsilon = 0.2$. 
	(b) In this simulation, we explore how $\widehat{M}(\epsilon)$ (for a fixed $\epsilon = 0.1$) varies with the $\mathcal{A}$-eigenvalues for the system defined in Figure \ref{fig:d1_3theta_STE}(a). We plot the theoretical values of $\widehat{M}(\epsilon)$ (given in \eqref{eq:def_widehatM}) for $\theta$ varying from $0$ to $\pi/2$ together with its actual values (as described in the text) obtained by running experiments for each $\theta$.
	}
	\label{fig:d3_Ae_Kne_Mhat_STE}
	\vspace{-8mm}
\end{figure*}

Given that Theorem~\ref{thm:stability_thm} seems to be correctly capturing the convergence speed dependence on $\nu$, the last facet of the problem to explore is the tightness of this bound.  Specifically, given a system of class $\mathcal{A}(d)$ and an observation function $h$, it is often of interest to estimate the minimum number of measurements $\left(\widehat{M}(\epsilon)\right)$ needed to ensure that for any $M' \ge M$ the conditioning number $\delta(M')$ is at most $\epsilon$ above the asymptotic level of $\delta_0$ (such an estimate is given in \eqref{eq:def_widehatM}).  To examine this, we refer back to the simulation shown in Figure~\ref{fig:edu_STE}(a) with parameters given in Table~\ref{tab:ex_d3_Ae_Kne}. Fixing $\epsilon = 0.2$, Figure~\ref{fig:d3_Ae_Kne_Mhat_STE}(a) re-plots $\max\{Q\}$ together with the line $C(1 + \delta_0 + \epsilon)$. 
Using the given parameters and \eqref{eq:def_widehatM} we  calculate that $\widehat{M}(\epsilon)\approx 166$.   Note that this value is also the intersection of the curve $C(1+ \delta(M))$ with the line $C(1+\delta_0 + \epsilon)$. 
Figure~\ref{fig:d3_Ae_Kne_Mhat_STE}(a) shows that $\max\{Q\}$ actually met this tolerance with only around 30 measurements. 
Thus, although the theoretical value of $\widehat{M}(\epsilon)$ given by \eqref{eq:def_widehatM} is correct, it is pessimistic in at least this particular case.

To demonstrate that the linear dependence of $\widehat{M}(\epsilon)$ on $\nu$ is correctly captured in the theorem, we restrict ourselves to $d = 1$.
Recall that when $d=1$, $\nu = |\sin(\theta)|^{-1}$ (since $T_s = 1$) where $\pm j\theta$ are the sole $\mathcal{A}$-eigenvalues.  
We repeat the simulation shown in Figure~\ref{fig:d1_3theta_STE}(a), this time using 100 values of $\theta$ equally spaced between $(0, \pi/2)$.  Fixing $\epsilon = 0.1$,  for each value of $\theta$ we note the value of $M$ where for all $M' > M$, $\max\left\{ \frac{\max\{Q\}}{C} - 1, \; 1 - \frac{\min\{Q\}}{C} \right\} < \delta_0 + \epsilon$. We call this value the ``actual'' $\widehat{M}(\epsilon)$, in contrast to the ``theoretical'' $\widehat{M}(\epsilon)$ given by \eqref{eq:def_widehatM}. Figure~\ref{fig:d3_Ae_Kne_Mhat_STE}(b) shows these actual and theoretical values of $\widehat{M}(\epsilon)$ as a function of $\theta$.  This comparison shows that while the theoretical $\widehat{M}(\epsilon)$ captures the same trend as the actual $\widehat{M}(\epsilon)$, the theoretical estimate can be pessimistic compared to the empirical values (though it is not clear if the theoretical bounds are achieved by some systems).

\subsection{Stable Embeddings for Dimension Estimation}
\label{sec:dimest}

To demonstrate the value of stable Takens' embeddings, this section will explore a simulated task estimating the dimensionality of an attractor.  The \emph{correlation dimension} is a measure of attractor dimension often applied to strange attractors of chaotic systems~\cite{grassberger1983measuring}, which corresponds to the actual geometric dimension of regular objects such as the circles and ellipses seen in linear system attractors~\cite{kantz2004nonlinear}.  To be precise, we first define 
the \emph{correlation sum} of tolerance $\epsilon$ for a set of 
points $\{x_k\}$ lying on a subset $\mathcal{M}$ and temporally related via the flow (i.e., $x_k = \Phi^k x_0$) as 
\begin{eqnarray}
	C(\epsilon, K) := \frac{2}{K(K-1)} \sum_{p = 1}^{K} \sum_{q = p+1}^{K} \Theta(\epsilon - \|F(x_p) - F(x_q)\|_2),
	\label{eq:corr_sum}
\end{eqnarray}
where $F$ is the delay coordinate map and $\Theta(\cdot)$ is the \emph{Heaviside step function} defined as $\Theta(x) = 0$ if $x \le 0$ and $\Theta(x) = 1$ if $x > 0$.
The \emph{correlation dimension} is defined  as $D = \lim_{\epsilon \rightarrow 0} \lim_{K \rightarrow \infty} \frac{\partial \log C(\epsilon, K)}{\partial \log \epsilon}$.
This makes intuitive sense as in the limit of small $\epsilon$ and large $K$, we expect $C(\epsilon,K)$ to scale like $C(\epsilon, K) \propto \epsilon^{-D}$, where $D$ is the dimension of the subset $\mathcal{M}$ in question.  Theoretically, one way to estimate correlation dimension is to plot the graph of $\log C(\epsilon, K)$ against $\log \epsilon$ for a large value of $K$, then simply read off the gradient for small values of $\log \epsilon$. 
In the absence of noise and with a topology preserving Takens' embedding  (i.e. $M > 2d$), this estimate should be as good as if one had access to the hidden system state. However, when noise is present, small values of $\log \epsilon$ will be capturing the noise characteristics and overestimating the attractor dimension.  A common approach in this case is to plot the local gradient $D(\epsilon) = \frac{\partial \log C(\epsilon,K)}{\partial \log \epsilon}$ against $\log (\epsilon)$ for a large value of $K$ and read off an estimate of the correlation dimension $D$ from a plateau in the graph, preferably in the regime of small $\epsilon$.

\begin{figure*}
	\hfil
	\begin{minipage}[t]{0.4\linewidth}
	\centerline{\epsfysize = 50mm \epsffile{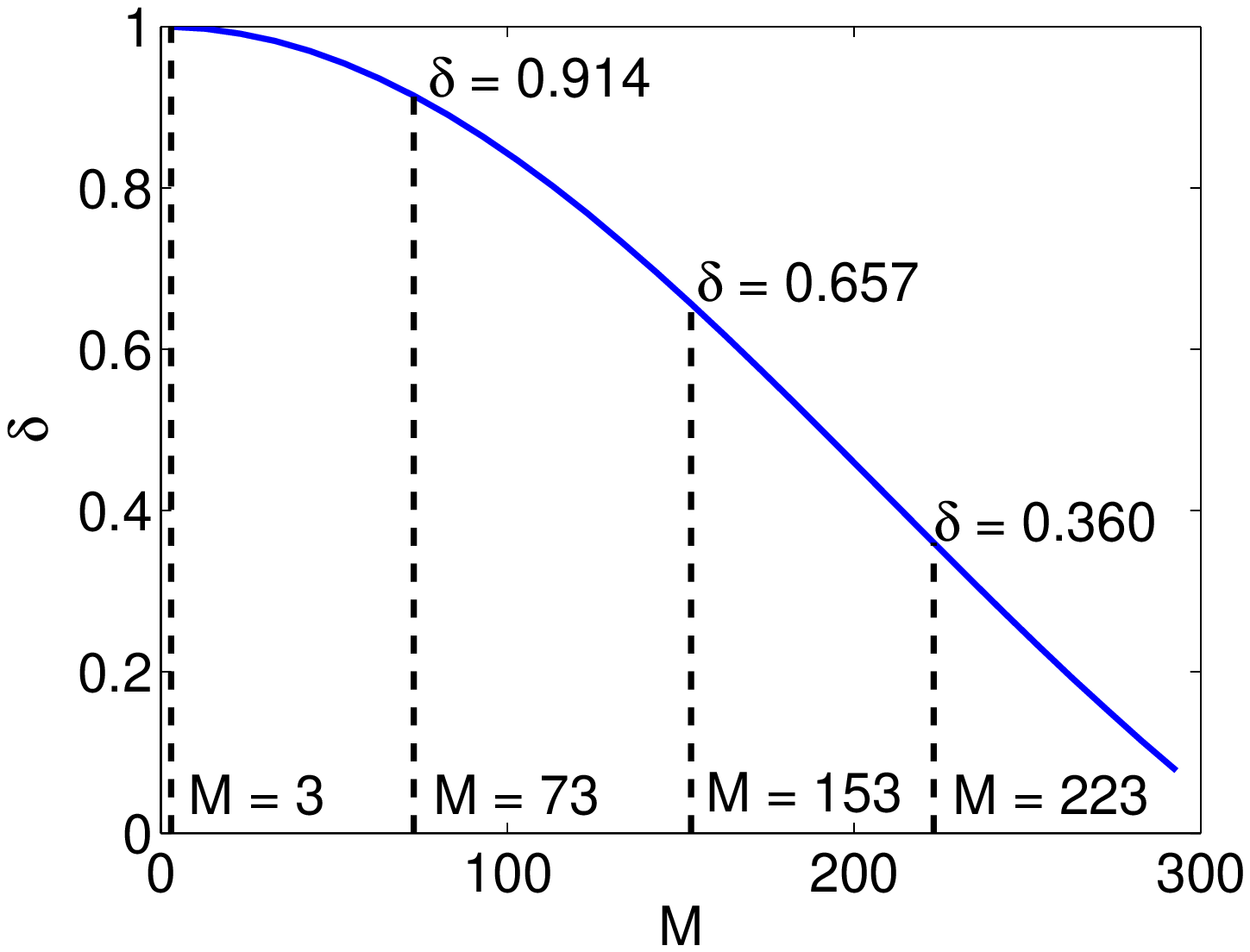}}
	\vspace{-1mm}
	\centerline{\small$\quad$~(a)}
	\end{minipage}
	\begin{minipage}[t]{0.4\linewidth}
	\centerline{\epsfysize = 50mm \epsffile{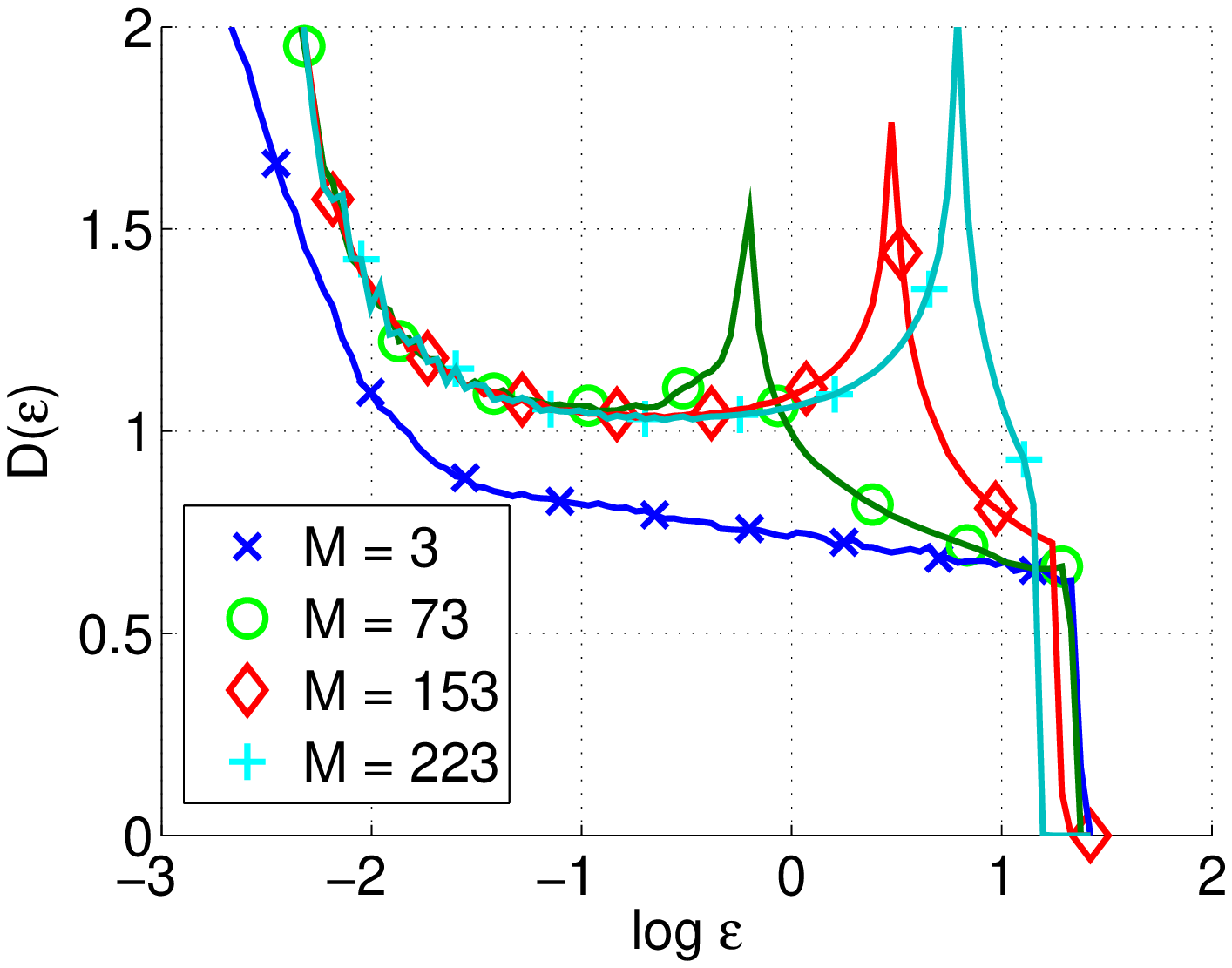}}
	\vspace{-1mm}
	\centerline{\small$\quad$~(b)}
	\end{minipage}
	\vspace{-1mm}
	\caption{\sl\small 
	Estimating the correlation dimension of a circular attractor $\mathcal{M}$ of a linear system of class $\mathcal{A}(1)$. (a) The conditioning of the stable embedding decreases with increasing number of measurements $M$. %
	(b) The graphs of $D(\epsilon)$ for the various $M$ considered are plotted against $\log \epsilon$. The correlation dimension estimate can be read off the plateaus in these graphs. 
	These plateau regions become more distinct with increasing $M$ (improving conditioning), and appear to converge to a value near the true dimension of 1.}
	\label{fig:corr_dim}
	\vspace{-8mm}
\end{figure*}

In this section, we use the above approach to estimate the correlation dimension of linear system attractors $\mathcal{M}$ in the reconstruction space $\reals^M$.  For this simulation construct a linear dynamical system of class $\mathcal{A}(1)$ with $N=100$, $\mathcal{A}$-eigenvalue $\theta = \frac{\pi}{300}$ and $\mathcal{A}$-eigenvector  $v = [1,\; j]^T$ (resulting in $A_1 = A_2$ and a circular attractor).  We also choose $h = [1, \; 1]^T$, implying that  $\kappa_1 = \kappa_2$ and subsequently $\delta_0=0$.
Figure \ref{fig:corr_dim}(a) shows that the actual conditioning\footnote{By actual conditioning, we mean the empirical value $\delta = \max\left\{ \frac{\max\{Q\}}{C} - 1, 1 - \frac{\min\{Q\}}{C} \right\}$, for $Q$ defined in Section \ref{sec:sims}.}
of $F$ approaches zero as we increase $M$.
To simulate noisy measurements, we corrupt the resulting time series formed by $h$ by adding white gaussian noise with zero mean and standard deviation $\sigma = 0.05$ (to give an SNR of about $32 dB$).

Figure \ref{fig:corr_dim}(b) shows the plots of $D(\epsilon)$ against $\log (\epsilon)$  with a number of delays $M = 3, 73, 153, 223$. 
For the graph corresponding to $M = 223$, a plateau is easily seen between $-1 < \log \epsilon < 0$, and corresponding to a correct dimension estimate of approximately 1.  We observe that by taking more measurements (i.e.,  improving the conditioning of the embedding), the estimate of the correlation dimension also improves.  Moreover, the width of the plateau region where we read off the correlation dimension estimate increases with increasing $M$, thus making its estimate more precise.
Note that when we take the minimum number of measurements $M = 3$ required by Takens' Theorem, there is no discernible plateau region in Figure \ref{fig:corr_dim}(b) for us to estimate the correlation dimension, and even the most reasonable estimate near $\log \epsilon=1$ is less accurate than with the estimates produced by the embeddings with better conditioning.

\section{Conclusion}

The main result of this paper has established that a delay coordinate map (using linear observation functions) can form a stable embedding for all pairs of points on the attractor of a linear dynamical system of class $\mathcal{A}(d)$.  The explicit, deterministic and non-asymptotic sufficient conditions we give for this stable embedding yield several observations about the embedding itself and favorable properties of system and measurement pairs.  For example, for many system and measurement pairs, the conditioning number $\delta(M)$ reaches a non-zero asymptotic value of $\delta_0$ with increasing $M$.   This ``plateau effect'' is in contrast with typical CS results where the conditioning of the stable embedding can be continually improved by increasing the number of measurements.  Furthermore, the convergence speed of the embedding quality to this limit is governed by the joint relationship of the system eigenvalues, which capture the relative speed with which the system explores the different dimensions of the state space (i.e., more diversity in these speeds results in faster convergence). Finally, we also see that the minimum number of delays $M$ of the delay coordinate map scales linearly with the attractor dimension but is independent of the system dimension. 
This is again in contrast with typical CS results, where the number of compressive measurements also scales logarithmically with the system dimension (but interestingly does parallel recent improvements in these bounds for the stable embedding of manifolds~\cite{clarkson2008tighter}).

While the comparisons with standard CS results reveal these interesting and non-intuitive technical differences between the results in each case, these discrepancies actually point to a much deeper difference in the problem setups that must be appreciated when embedding attractors of dynamical systems.  Perhaps the easiest way to see this is to consider that in the present case of delay coordinate maps, while the number of measurements doesn't scale with the ambient system dimension, the total number of measurements may in fact have to be larger than the system dimension ($M>N$) in order to make a particular  conditioning guarantee.  In the typical CS case, this would of course be a ridiculous proposition.  If the RIP property required ($M>N$) random measurements (e.g., due to very large constants in the typical sufficient conditions), one would likely abandon the CS strategy and simply take $N$ uncoded measurements (e.g., in the canonical basis).  However, in the case of delay coordinate maps for dynamical systems, this luxury is simply not available.  For example, the observers often do not have any control over the choice of observation function $h$, and in these cases cannot simply change the way the system is measured.  But, more importantly, even if we were given complete control over $h$, it is only a ``seed'' that is used in producing the whole measurement process.  One can view the entire set of measurements as being generated by repeatedly forcing this observation function through the dynamics of the system (seen explicitly in writing the delay coordinate map in~\eqref{eq:DCM_linear_arb_h}).  Said another way, because there is only a single observation function for the system, the total measurement process for a delay coordinate map is beholden to the dynamics of the system itself to provide sufficient diversity to make the measurements informative.  Therefore, even with complete control over the observation function, delay coordinate maps represent a highly restricted total measurement process that cannot be completely controlled (without access to and control over the system that is hidden and in need of measurement).

Characterizing the delay coordinate map embeddings for attractors of linear dynamical systems with linear observation functions is a subset of the more general problem of characterizing these embeddings for attractors of nonlinear systems and general observation functions.  
From the results here, we conclude that there is reason to be optimistic that similar stability results can be obtained for this more general case of interest.  Furthermore, these results also lead us to conclude that there are several issues that differ from standard CS results and will need to be carefully considered in any generalization.  

\appendices

\section{Proof of Stable Takens' Embedding Theorem}

\label{sec:proof_main}

Because Theorems \ref{thm:existence_thm} and \ref{thm:stability_thm} are very similar in structure, we will essentially lay out the proof approach for both of them together in this section and then separately establish the necessary details for each result.  Before proceeding with the specific proofs, we will introduce some notation and preliminary results that will be useful.

\subsection{Notation and preliminaries}

\subsubsection{Frame theory}

Drawing on some terminology from the field of \emph{frame theory}, we say that a sequence of vectors $\{g_i\}_{i=1}^{M}$ in $\comps^K$, $M \ge K$, forms a \emph{frame}~\cite{christensen2003introduction} for $\comps^K$ if there exists two real constants $0 < B_1 \le B_2 < \infty$ such that for all $\alpha \in \comps^K$, $B_1\|\alpha\|_2^2 \le \sum_{i=1}^{M} |\langle g_i,\alpha \rangle|^2 = \|G \alpha\|_2^2 \le B_2 \|\alpha\|_2^2,$ where $G^H = \left(g_1 \; | \; g_2 \; | \; \cdots \; | \; g_M\right) \in \comps^{K \times M}$, the concatenation of the $\{g_i\}_{i=1}^{M}$, is called the \emph{frame analysis operator} and $B_1, B_2$ are called the \emph{frame bounds}.  The frame bounds can be defined as $B_1 = \lambda_{\min}$ and $B_2 = \lambda_{\max}$, where $\lambda_{\min}$ and $\lambda_{\max}$ are the minimum and maximum eigenvalues of $G^H G \in \comps^{K \times K}$.

\subsubsection{Linear delay coordinate maps}

Because the attractor $\mathcal{M}$ is contained in the span of the columns of $V$, for any $x,y \in \mathcal{M}$ we can write $x = V \alpha_x$ and $y = V \alpha_y$ for some complex coefficients $\alpha_x, \alpha_y \in \comps^{2d}$. Using $F$ to denote the delay coordinate map for a linear system with flow matrix $\Phi$ and observation function $h$ as described in \eqref{eq:DCM_linear_arb_h}, the $k$-th row (for $k = 1, \cdots, M$) of the vector $F(x) - F(y)$
 can be written
	$h^T\left(\Phi^{k-1}(x-y)\right) 
	= h^T \left( \Phi^{k-1} V(\alpha_x - \alpha_y) \right)
	= h^T \left( V D^{k-1}(\alpha_x - \alpha_y) \right) = \langle g_k, \alpha_x - \alpha_y \rangle$,
where 
\begin{equation}
	g_k^H \hspace{-1mm}=  h^T V D^{k-1} = \left[ (v_1^T h) e^{\mbox{-}j(k-1)\theta_1 T_s } , (v_1^H h) e^{j(k-1)\theta_1 T_s } ,  \dots, (v_d^T h) e^{\mbox{-}j(k-1)\theta_d T_s}, (v_d^H h) e^{j(k-1)\theta_d T_s }  \right]	
	\label{eqn:gvec}
\end{equation}
and $D$ is the diagonal matrix comprised of $\mathcal{A}_{\Phi}$-eigenvalues as defined in Section \ref{sec:linear_sys}.
Thus, we have:
	$\|F(x) - F(y)\|_2^2 = \sum_{k = 1}^{M} |\langle g_k , (\alpha_x - \alpha_y) \rangle|^2
	=\|G(\alpha_x - \alpha_y)\|_2^2$,
where $G \in \comps^{M \times 2d}$ is the concatenation of $\{g_k\}$ as described above. In this following, $G$ is fixed to be this matrix given here.

\subsubsection{Eigenvalue bounds}
It will be important in the following proofs to determine bounds on the extreme eigenvalues of the matrix $G^H G$.  To that end, we first introduce the well-known
Gershgorin Circle Theorem, which we state here for notational convenience:
\begin{thm}[Gershgorin Circle Theorem \cite{moon2000mathematical}]
	\label{thm:Gershgorin_disk}
	The eigenvalues of a $K \times K$ matrix $A$ all lie in the union of the Gershgorin disks of $A$. The Gershgorin disk $\mathcal{D}_i$ for $i = 1, \cdots, K$, is defined as
		$\mathcal{D}_i = \left\{ x \in \comps\;:\; |x-\mathcal{C}_i| \le \widetilde{r}_i \right\},$
	where $\widetilde{r}_i := \sum_{j=1,\;j\neq i}^{K} |(A)_{i,j}|$ is the radius, and $\mathcal{C}_i := (A)_{i,i}$ is the center of the $i$-th  disk.
	Thus
		$\lambda(A) \subset \bigcup_{i=1}^{K} \mathcal{D}_i,$
	where $\lambda(A) = \{ \lambda_1, \cdots, \lambda_K \}$, and $\{\lambda_i\}$ are the eigenvalues of $A$.
\end{thm}

To apply the Gershgorin Circle Theorem  to obtain the extrema eigenvalues of $G^H G$, we introduce the following useful lemma that gives values for centers $\mathcal{C}_i$ and radii $\widetilde{r}_i$ of the Gershgorin disks $\mathcal{D}_i$ of $G^H G$.
\begin{lemma}
	\label{lem:Gershgorin_disk}
	For $i = 1, \cdots, d$, the centers of the Gershgorin disks of $G^H G$ are $\mathcal{C}_{2i-1} = \mathcal{C}_{2i} = |v_i^H h|^2 M$ while their radii are 
		$\widetilde{r}_{2i-1} = \widetilde{r}_{2i} = |v_i^H h|^2 \left| \frac{\sin (M\theta_iT_s)}{\sin (\theta_i T_s)} \right| + \sum_{p=1, \;p\neq i}^{d} |v_i^H h||v_p^H h| \left| \frac{\sin \left( M(\theta_i - \theta_p)T_s/2 \right)}{\sin \left( {(\theta_i - \theta_p)T_s/2} \right)} \right|  
		+\sum_{p=1, \;p\neq i}^{d} |v_i^H h||v_p^H h| \left| \frac{\sin \left( {M(\theta_i + \theta_p)T_s/2} \right)}{\sin \left( {(\theta_i + \theta_p)T_s/2} \right)} \right|$.
\end{lemma}

\begin{proof}
	We can write $G^H G = \sum_{k=1}^{M} g_k g_k^H,$ 
	where we recall that $g_k$ is defined as in \eqref{eqn:gvec}.
	Thus the $(p,q)$ entry of $G^H G$ can be expressed as: $(G^H G)_{p,q} = \sum_{k=1}^{M} g_k(p) g_k(q)^{*}$, where $g_k(p)$ denotes the $p$-th entry of the vector $g_k$.
	As such, the formation of $G^H G$ involves the calculation of sum of complex trigonometric polynomials due to the complex exponentials $\left(\{e^{\pm j (k-1) \theta_p T_s} \}\right)$ appearing in the terms of each $g_k$.  A few separate cases need to be considered because of the differences in the even ($2p$) and odd ($2p-1$) numbered rows of $G^H G$ for all $p$.  We first consider the even numbered rows.  The diagonal terms actually have a fairly simple form:
	$(G^H G)_{2p,2p} = \sum_{k=1}^{M} g_k(2p) g_k(2p)^{*}=\sum_{k=1}^{M} |v_p^H h|^2 = M |v_p^H h|^2$.
	The adjacent term to the left is given by:
	$(G^H G)_{2p,2p-1} = 
	\sum_{k=0}^{M-1} \left( (v_p^T h) e^{-j k \theta_p T_s} \right)^2 =
	(v_p^T h)^2 \sum_{k=0}^{M-1} \left(e^{-j 2\theta_p T_s}\right)^k =
	(v_p^T h)^2 \frac{\sin(M\theta_p T_s)}{\sin(\theta_p T_s)}e^{-j(M-1)\theta_p T_s}$,
	where the last expression follows from the standard formula for a finite geometric sum, pulling out common exponential factors, and using Euler's formula.  The other cross terms for all $p,q \in \{1,\dots, d\}$ such that $p \neq q$ can be derived similarly as:
	\begin{eqnarray*}
		(G^H G)_{2p,2q} &= (v_p^T h)(v_q^H h) \sum_{k=0}^{M-1} \left( e^{-j 2 \left( \frac{\theta_p - \theta_q}{2} \right) T_s} \right)^k
		= (v_p^T h)(v_q^H h) \frac{\sin\left(M\left( \frac{\theta_p - \theta_q}{2} \right) T_s\right)}{\sin\left(\left( \frac{\theta_p - \theta_q}{2} \right) T_s\right)} e^{-j(M-1)\left( \frac{\theta_p - \theta_q}{2} \right) T_s},\\
		(G^H G)_{2p,2q-1} &= (v_p^T h)(v_q^T h) \sum_{k=0}^{M-1} \left( e^{-j 2 \left( \frac{\theta_p + \theta_q}{2} \right) T_s} \right)^k
		= (v_p^T h)(v_q^T h) \frac{\sin\left(M\left( \frac{\theta_p + \theta_q}{2} \right) T_s\right)}{\sin\left(\left( \frac{\theta_p + \theta_q}{2} \right) T_s\right)} e^{-j(M-1)\left( \frac{\theta_p + \theta_q}{2} \right) T_s}.
	\end{eqnarray*}
	The relevant quantities for the odd numbered rows are given similarly as
	\begin{align*}
		(G^H G)_{2p-1,2p-1} & = (G^H G)_{2p,2p} =M |v_p^H h|^2,\\
		(G^H G)_{2p-1,2p} &= 
		 (G^H G)_{2p,2p-1}^{*} = (v_p^H h)^2 \frac{\sin(M\theta_p T_s)}{\sin(\theta_p T_s)}e^{j(M-1)\theta_p T_s}
		,\\
		(G^H G)_{2p-1,2q} &= 
		(G^H G)_{2q,2p-1}^{*} = (v_q^H h)(v_p^H h) \frac{\sin\left(M (\theta_q + \theta_p)  T_s/2 \right) }{\sin\left((\theta_q + \theta_p) T_s/2 \right)}e^{j(M-1)\left( \frac{\theta_q + \theta_p}{2} \right) T_s}
		,\\
		(G^H G)_{2p-1,2q-1} &= 
		(v_p^H h)(v_q^T h) \frac{\sin\left(M(\theta_p - \theta_q) T_s/{2} \right)}{\sin\left( (\theta_p - \theta_q) T_s/{2} \right)}e^{j(M-1)\left( \frac{\theta_p - \theta_q}{2} \right) T_s}.
	\end{align*}
	Finally we note that many of the above complex quantities only differ in their phase because of symmetry in the summations, making their magnitudes equal when calculating the radii of the Gershgorin disks.  The expressions for $\mathcal{C}_i$ and $\widetilde{r}_i$ in the lemma are obtained simply by applying the notation of the Gershgorin Circle Theorem to the calculated magnitudes of the entries of $G^H G$.
\end{proof}

\subsection{General proof approach}
\label{sec:approach}
Using the preliminaries above, we can now sketch out the general approach for the proof of both theorems below.  Essentially, the theorems result from using (or establishing) the following three facts:

\begin{enumerate}
	\item If $G^H G \in \comps^{2d \times 2d}$ is established to be full rank, then $\{g_k\}_{k=1}^{M}$ form a frame in $\comps^{2d}$. Thus there exists $0 < B_1 \le B_2 < \infty$ such that
		$B_1 \le \frac{\|F(x) - F(y)\|_2^2}{\|\alpha_x - \alpha_y\|_2^2} \le B_2$  
	holds for all distinct pairs of points $x,y \in \mathcal{M}$.
	In particular, to establish conditioning guarantees, we can let $B_1$ and $B_2$ be the smallest and largest eigenvalues of $G^H G$ (respectively) and determine bounds on those important quantities.

	\item Next, we use the fact that $\|x-y\|_2^2 = (\alpha_x - \alpha_y)^H V^H V (\alpha_x - \alpha_y)$ to get
		$A_1 \le \frac{\|x-y\|_2^2}{\|\alpha_x - \alpha_y\|_2^2} \le A_2$, 
	where $A_1$ and $A_2$ are the smallest and largest eigenvalues of $V^H V \in \comps^{2d \times 2d}$ respectively. By the definition of $V$ we know that $V^H V$ is well-defined and full rank, meaning that $0 < A_1 \le A_2 < \infty$.
	
	\item Putting the 2 previous steps together, we get 
		$0 < \frac{B_1}{A_2} \le \frac{\|F(x) - F(y)\|_2^2}{\|x-y\|_2^2} \le  \frac{B_2}{A_1} < \infty,$
	where the bounds $\frac{B_1}{A_2}$ and $\frac{B_2}{A_1}$ can be manipulated to get the scaling constant $C$ and conditioning $\delta$ in \eqref{eqn:ARIP}.  Specifically, we can set $C = \frac{1}{2}\left(\frac{B_1}{ A_2} + \frac{B_2}{ A_1} \right)$ and $\delta~=~1~-~\frac{B_1}{ C A_2}$.
\end{enumerate}
		
	\subsection{Proof of Theorem \ref{thm:existence_thm}}
\begin{proof}
	For Theorem \ref{thm:existence_thm}, we follow the three steps detailed in Appendix~\ref{sec:approach}, where we only need to show that $G^H G$ is indeed full rank given the conditions of the theorem.
	Consider first the case when $M=2d$, where showing $G^H G$ is full rank is equivalent to showing $\det(G^H G) = \det(G)^2 > 0$. 
	The matrix $G$ can be expressed in terms of a product of a Vandermonde matrix and a diagonal matrix:
	\begin{eqnarray*}
		G 
		&=& \left( 
		\begin{smallmatrix}
			1 & 1 & \cdots & 1 & 1 \\
			e^{-j\theta_1 T_s} & e^{j\theta_1 T_s} & \cdots & e^{-j\theta_d T_s} & e^{j\theta_d T_s} \\
			\vdots & \vdots &  & \vdots & \vdots \\
			e^{-j2d \theta_1 T_s} & e^{j2d \theta_1 T_s} & \cdots & e^{-j2d \theta_d T_s} & e^{j2d \theta_d T_s} \\
		\end{smallmatrix}
		\right)
		\left( 
		\begin{smallmatrix}
			v_1^T h & & & & (0) \\
			& v_1^H h & & & \\
			& & \ddots & & \\
			& & & v_d^T h & \\
			(0) & & & & v_d^H h
		\end{smallmatrix}
		\right) 
		= \widetilde{M}^T \widetilde{H},  \nonumber
	\end{eqnarray*}
	where $\widetilde{M}$ is the Vandermonde matrix with 
	the $\mathcal{A}_{\Phi}$-eigenvalues as its parameters and $\widetilde{H}$ is a diagonal matrix whose diagonal elements are made up of the projection of $h$ onto the $\mathcal{A}$-eigenvectors.
	Thus, $\det(G) = \det(\widetilde{M}) \det(\widetilde{H})$. 
	One of the conditions of Theorem~\ref{thm:existence_thm} ensures that the $\{e^{\pm j \theta_i T_s}\}_{i=1}^{d}$ are distinct, which implies that the determinant of this square Vandermonde matrix \cite[Ch 0]{horn1990matrix} obeys $|\det(\widetilde{M})| > 0$. Also since $v_i^H h \neq 0$ for all $i = 1, \cdots, d$, we also know that $|\det(\widetilde{H})| > 0$. Therefore for $M = 2d$, $\operatorname{rank}(G^H G) = 2d$. Since adding vectors to a frame does not change the rank of $G^H G$ (i.e., frame bounds cannot be lowered by adding more vectors to the frame), it follows that if $M \ge 2d$ then	$\operatorname{rank}(G^H G) = 2d$ and the proof of Theorem~\ref{thm:existence_thm} is complete.
	\end{proof}

	\subsection{Proof of Theorem \ref{thm:stability_thm}}
	\label{sec:proof_of_stability_thm}
	\begin{proof}
	To prove Theorem \ref{thm:stability_thm}, we again follow the three steps detailed in Appendix~\ref{sec:approach}, this time establishing specific guarantees on the frame bounds $B_1(M)$ and $B_2(M)$ appearing in the first step. 
	From Lemma~\ref{lem:Gershgorin_disk}, we first observe that for all $i$ we can bound the Gershgorin disk radii by 
	$\widetilde{r}_{2i-1} = \widetilde{r}_{2i} \le 
			\left(|v_i^H h|^2   + \sum_{p=1,\;p\neq i}^{d} |v_i^H h||v_p^H h|  + 
			\sum_{p=1,\; p\neq i}^{d} |v_i^H h||v_p^H h|\right) \nu \leq (2d-1)\kappa_2^2 \|h\|_2^2 \nu$. 
		Noting that $\|h\|_2^2 = \frac{2d}{M}$, we see that for each $i$, the Gershgorin disks of $G^H G$ satisfy $\mathcal{D}_{2i-1} = \mathcal{D}_{2i} \subset \left[
		|v_i^H h|^2 M - \|h\|_2^2 (2d - 1) \nu \kappa_2^2, \; |v_i^H h|^2 M + \|h\|_2^2 (2d - 1) \nu \kappa_2^2
		\right]$.
		Then applying the Gershgorin Circle Theorem, we get $\lambda(G^H G) \subset \bigcup_j^{2d} \mathcal{D}_j \subset \left[ 2d \kappa_1^2  - \frac{2d}{M} (2d - 1) \nu \kappa_2^2,\; 2d \kappa_2^2 + \frac{2d}{M} (2d - 1) \nu \kappa_2^2 \right]$.
		By choosing $B_1(M) = 2d \left( \kappa_1^2 - \frac{(2d-1)\nu \kappa_2^2}{M} \right)$ and $B_2(M) = 2d \left( \kappa_2^2 + \frac{(2d-1)\nu \kappa_2^2}{M} \right)$, and applying step 2 in Section \ref{sec:approach},  %
		we arrive at:
		\begin{eqnarray}
			\frac{B_1(M)}{A_2} \le \frac{\|F(x) - F(y)\|_2^2}{\|x-y\|_2^2} \le  \frac{B_2(M)}{A_1}
			\label{eq:iso_bounds_repeat}
		\end{eqnarray}
		for all distinct pairs of points $x,y \in \mathcal{M}$ and for all $M$. 
		
		Now as $M \rightarrow \infty$, $B_1(M) \rightarrow 2d \kappa_1^2$ and $B_2(M) \rightarrow 2d \kappa_2^2$. Thus in the limit of large $M$, the lower and upper bounds of the inequality \eqref{eq:iso_bounds_repeat} approaches $\frac{2d \kappa_1^2}{A_2}$ and $\frac{2d \kappa_2^2}{A_1}$, respectively. We define the scaling constant $C$ as the average of the asymptotic values of these lower and upper bounds: $C := \frac{2d}{2}\left(\frac{\kappa_1^2}{A_2} + \frac{\kappa_2^2}{A_1} \right)$. Also define the conditioning number $\delta(M)$ for a given $M$ as the maximum deviation of the lower and upper bounds of \eqref{eq:iso_bounds_repeat} from $C$, normalized by $C$:
			$\delta(M) := \max \left\{ 1 - \frac{B_1(M)}{C A_2}, \frac{B_2(M)}{C A_1} - 1 \right\}.$
		Now $1 - \frac{B_1(M)}{C A_2} =  1 - \frac{2d \left( \kappa_1^2 - {(2d-1)\nu \kappa_2^2/M} \right)}{(2d/2)\left({\kappa_1^2} + \kappa_2^2(A_2/A_1) \right)}
		= \frac{A_2 \kappa_2^2 - A_1 \kappa_1^2 + {2 A_1 (2d-1)\nu \kappa_2^2/M}}{A_2 \kappa_2^2 + A_1 \kappa_1^2}$, and 
		$\frac{B_2(M)}{C A_1} - 1 =  \frac{2d \left( \kappa_2^2 + {(2d-1)\nu \kappa_2^2/M} \right)}{(2d/2)\left((A_1/A_2){\kappa_1^2} + \kappa_2^2 \right)} - 1
		= \frac{A_2 \kappa_2^2 - A_1 \kappa_1^2 + {2 A_2 (2d-1)\nu \kappa_2^2/M}}{A_2 \kappa_2^2 + A_1 \kappa_1^2}$.
		Since $A_1 \le A_2$, we have that
		$\delta(M) = \frac{B_2(M)}{C A_1} - 1 = \frac{A_2 \kappa_2^2 - A_1 \kappa_1^2}{A_2 \kappa_2^2 + A_1 \kappa_1^2} + \frac{2 A_2 \kappa_2^2}{A_2 \kappa_2^2 + A_1 \kappa_1^2}\frac{(2d-1)\nu}{M}.$
		We can then define $\delta_0$ and $\delta_1(M)$ as the first and second term of the sum above. Notice that $\delta(M)$ represents a worst case bound on the deviation from $C$, as we maximized over upper and lower bounds that may not be the same magnitude (i.e., in general $C(1 - \delta(M)) \neq \frac{B_1(M)}{A_2}$).

		Finally, we recall that for the embedding conditioning number to be valid, we must have $0 \le \delta(M) < 1$. The first condition $\delta(M) \ge 0$ is achieved by construction. %
		The upper bound is equivalent to the condition for $M$ required by the theorem statement, thus completing the proof. 
\end{proof}		

\section*{Acknowledgment}

The authors are grateful to Michael Wakin and Armin Eftekhari for valuable discussions about this work.

\bibliographystyle{IEEEtran}
\bibliography{IEEEabrv,refs}



\end{document}